\documentclass{article}
\usepackage{arxiv}
\usepackage[ruled,linesnumbered,onelanguage]{algorithm2e}
\usepackage{graphicx}
\usepackage{hyperref}
\usepackage{listings}
\usepackage[dvipsnames]{xcolor}
\usepackage{subcaption}
\usepackage{times}
\usepackage{todonotes}
\usepackage{url} 
\usepackage{multirow}  
\usepackage{colortbl}
\usepackage{wrapfig}
\usepackage{floatflt}
\usepackage{float}
\usepackage{amsmath}

\usepackage{amsthm}
\usepackage{amsfonts}
\usepackage{booktabs}
\usepackage[binary-units=true]{siunitx}

\usepackage{amssymb}
\usepackage{pifont}
\usepackage{balance}

\newcommand{\secref}[1]{Section~\ref{#1}}
\newcommand{\blue}[1]{\textcolor{blue}{#1}}
\newcommand{\green}[1]{\textcolor{olive}{#1}}
\newcommand{\red}[1]{\textcolor{red}{#1}}

\definecolor{aogreen}{rgb}{0.0, 0.5, 0.0}
\newcommand{\dgreen}[1]{\textcolor{aogreen}{#1}}

\newcommand{\ttt}[1]{\texttt{#1}}

\definecolor{white}{rgb}{1,1,1}
\definecolor{backgroundColour}{rgb}{0.95,0.95,0.92}
\definecolor{mGray}{rgb}{0.5,0.5,0.5}
\definecolor{greencomment}{rgb}{0.247,0.498,0.372}

\lstdefinestyle{JavaStyle}{
    backgroundcolor=\color{backgroundColour}, 
  language=Java,
  tabsize=2,
  breaklines=true,
  breakatwhitespace=true,
  escapechar=|*,
  numbers=left,                    
  numbersep=5pt,
  stepnumber=1,
  basicstyle=\fontsize{8.5pt}{10pt}\ttfamily,
  numberstyle=\footnotesize\ttfamily\color{mGray},
  aboveskip=\baselineskip,
  captionpos=b,
  frame=single,
  columns=fullflexible,
  showstringspaces=false,
  extendedchars=true,
  breaklines=true,
  showtabs=false,
  showspaces=false,
  identifierstyle=\ttfamily,
  keywordstyle=\color[rgb]{0.498,0.0,0.333},
  stringstyle=\color[rgb]{0.165,0.0,0.999},
  commentstyle=\color[rgb]{0.247,0.498,0.372},
  morekeywords={String}
}

\lstdefinestyle{UseCase}{
	style=JavaStyle,
	captionpos=b,
	numbers=none,
	float=h,
	frame=none,
	backgroundcolor=\color{white},
 tabsize=100
}

\theoremstyle{definition}
\newtheorem{example}{Example}[section]

\newtheorem{definition}{Definition}[section]
\newcommand\tuple[1]{\langle #1 \rangle}
\newtheorem{remark}{Remark}

\newcommand{\defas}[0]{\stackrel{{\scriptscriptstyle\rm def}}{=}}

\newcommand{\inlineJava}[1]{\lstinline[style=UseCase,basicstyle=\fontsize{10pt}{12pt}\ttfamily,breaklines=true]!#1!}

\newcommand{\locator}{selector} 
\newcommand{\locators}{selectors} 
\newcommand{\Locators}{Selectors}

\newcommand{\Methods}{{\mathit{Methods}}} 
\newcommand{\Blocks}{{\mathit{Blocks}}} 
\newcommand{\Instrs}{{\mathit{Instrs}}} 
\newcommand{\Shadows}{{\mathit{Shadows}}}

\newcommand{\blocks}{{\mathrm{blocks}}} 
\newcommand{\instrs}{{\mathrm{instrs}}}

\newcommand{\shadowequiv}{\equiv_{\mathrm{s}}^{m}}

\newcommand{\cmark}{\ding{51}}%
\newcommand{\xmark}{\ding{55}}%
\newcommand{\both}[1]{\multirow{2}{*}{#1}}
\newcommand{\hmark}{\cmark\kern-1.1ex\protect\raisebox{.7ex}{\protect\rotatebox[origin=c]{125}{--}}}%

\begin{document}
\title{Efficient and Expressive Bytecode-Level Instrumentation for Java Programs}
\author{
  Chukri~Soueidi\\
  Univ.~Grenoble~Alpes,~Inria,\\ 
  CNRS,~Grenoble~INP,~LIG\\ 
  38000~Grenoble\\ 
  France \\
  \texttt{chukri.a.soueidi@inria.fr} \\
  \And
  Marius~Monnier\\
  Univ.~Grenoble~Alpes,~Inria,\\ 
  CNRS,~Grenoble~INP,~LIG\\ 
  38000~Grenoble\\ 
  France \\
  \texttt{marius.monnier@inria.fr} \\
  \And
  Ali~Kassem\\
  Univ.~Grenoble~Alpes,~Inria,\\ 
  CNRS,~Grenoble~INP,~LIG\\ 
  38000~Grenoble\\ 
  France \\
  \texttt{ali.kassem@inria.fr} \\
  \And
  Yli\`{e}s~Falcone\\
  Univ.~Grenoble~Alpes,~Inria,\\ 
  CNRS,~Grenoble~INP,~LIG\\ 
  38000~Grenoble\\ 
  France \\
  \texttt{ylies.falcone@inria.fr} \\
}
    
%

\maketitle              
\begin{abstract}
We present an efficient and expressive tool for the instrumentation of Java programs at the bytecode-level.
BISM (Bytecode-Level Instrumentation for Software Monitoring) is a light\-weight Java bytecode instrumentation tool that features an expressive high-level control-flow-aware instrumentation language. 
The language is inspired by the aspect-oriented programming paradigm in modularizing instrumentation into separate \textit {transformers}, that encapsulate joinpoint selection and advice inlining.
BISM allows capturing joinpoints ranging from bytecode instructions to methods execution and provides comprehensive static and dynamic context information. 
It runs in two instrumentation modes: build-time and load-time. 
BISM also provides a mechanism to compose transformers and automatically detect their collision in the base program.
Transformers in a composition can control the visibility of their advice and other instructions from the base program.
We show several example applications for BISM and demonstrate its effectiveness using three experiments: a security scenario, a financial transaction system, and a general runtime verification case. 
The results show that BISM instrumentation incurs low runtime and memory overheads. 

\keywords{
Instrumentation \and
Runtime Verification \and
Monitoring \and
Java \and
Bytecode \and 
Aspect-Oriented Programming \and 
Control Flow \and 
Static and Dynamic Contexts
}
\end{abstract}  
\section{Introduction}

Instrumentation is essential in software engineering and verification activities.
In the software monitoring workflow~\cite{BartocciFFR18}, instrumentation allows extracting information from a running software to abstract the execution into a trace fed to a monitor.
Depending on the information needed by the monitor, the granularity level of captured and extracted information may range from coarse (e.g., a function call) to fine (e.g., an assignment to a local function/method variable, a jump in the control flow).

General-purpose tools for instrumenting Java programs have two important aspects that distinguish their usability: their level of expressiveness and level of abstraction.
Expressiveness determines how much the user can extract information from the bytecode and alter the program's execution.
Moreover, abstraction determines how much the user has to deal with bytecode details and how high-level the code has to be written.
Extracting fine-grained events requires tools with high expressiveness.
However, these tools are often too low-level and require expertise on the bytecode, making the instrumentation too verbose and error-prone.
Another requirement of an instrumentation process or tool is to have a low-performance impact on the target program, 
which can be measured by the runtime overhead of the inserted code, the memory overhead, and the size of the generated executable.

Aspect-oriented programming (AOP)~\cite{KiczalesLMMLLI97} is a popular and convenient paradigm for instrumenting a program. AOP advocates the modularization of crosscutting concerns such as instrumentation.
For Java programs, runtime verification tools \cite{FalconeKRT18,BartocciFBCDHJK19} have long relied on AspectJ~\cite{KiczalesHHKPG01}, which is one of the reference AOP implementations for Java.
At the core of AOP is the joinpoint model, which consists of joinpoints (points in the execution of a program) and pointcuts (a mechanism to select joinpoints).
AspectJ provides a high-level joinpoint model for convenient instrumentation.
Although AspectJ provides several functionalities oriented on the development process and program structure, these are seldom used in the context of verification.
AspectJ does not offer enough flexibility to capture and extract fine-grained information from the program.
This makes instrumentation tasks that require capturing low-level bytecode regions, such as bytecode instructions, local variables of a method, and basic blocks in the control-flow graph (CFG) unachievable with AspectJ. 
In \secref{sec:aes}, we demonstrate an instrumentation scenario that cannot be achieved with AspectJ.
Moreover, AspectJ provides limited support for writing program static analyzers that can be combined with runtime verification.
In particular, the provided static crosscutting constructs are only limited to inter-type declarations, weave-time error, and warning declarations, and exception softening.
These constructs are not expressive enough to allow users to define complex compile-time computations.

Nevertheless, there are several low-level bytecode manipulation frameworks such as ASM~\cite{BrunetonASM02} (in reference to the \lstinline|__asm__| C++ operator) and BCEL~\cite{BCEL} (Byte Code Engineering Library) which are highly efficient and expressive. 
%
However, writing instrumentation in such frameworks is too verbose, tedious, and requires expertise in bytecode. 
Other bytecode instrumentation frameworks, from which DiSL~\cite{MarekVZABQ12} is the most remarkable, enable flexible low-level instrumentation and, at the same time, provide a high-level language. 
However, DiSL does not support inserting bytecode instructions directly but allows writing custom transformers that allow the user to traverse the bytecode freely and modify it.
These custom transformers can be run only before the main DiSL instrumentation process, and the developer needs to revert to low-level bytecode manipulation frameworks to implement them.
This makes several scenarios tedious to implement in DiSL and requires workarounds that incur a considerable bytecode overhead.
For example, in \secref{sec:aes}, we demonstrate the overhead incurred by DiSL when instrumenting an inline monitor that duplicates if-state\-ments in a program.

\paragraph{Contributions.}
We introduce BISM (Bytecode-Level Instrumentation for Software Monitoring), a lightweight bytecode instrumentation tool for Java programs that features an expressive high-level instrumentation language. 
The language is inspired by AOP but adopts an instrumentation model that is more directed towards runtime verification.
In particular, BISM provides separate classes, \textit{transformers}, that encapsulate joinpoint selection and advice inlining.
It offers various instrumentation \emph{\locators{}} that select joinpoints covering bytecode instructions, basic blocks, and methods execution.
BISM also provides access to a set of comprehensive joinpoint-related \emph{static} and \emph{dynamic contexts} to retrieve relevant information.
BISM provides a set of \emph{advice methods} that specify the advice to insert code, invoke methods, and print information.

BISM is control-flow aware.
That is, it generates CFGs for all methods and provides access to them in its language. 
Moreover, BISM provides several control-flow properties, such as capturing conditional jump branches and retrieving successor and predecessor basic blocks. 
Such features can provide support to tools relying on a control-flow analysis, for instance, in the security domain, to check for control-flow integrity. 
BISM also provides a mechanism to compose multiple transformers and automatically detect their collision in the base program.
Transformers in a composition are capable of controlling the visibility of their advice and of other original instructions.
BISM is a standalone tool implemented in Java that requires no installation, as compared to AspectJ and DiSL.
BISM can run in two instrumentation modes: build-time to statically instrument a Java class or Jar file, and load-time to run as a Java agent that instruments classes being loaded by a running JVM.
We show several applications for BISM in static and dynamic analysis of programs.
We also demonstrate BISM effectiveness and performance using three complementary experiments.
The first experiment shows how BISM can be used to instrument a program to detect test inversion attacks.
For this purpose, we use BISM to instrument AES (Advanced Encryption Standard). 
The second and third experiments demonstrate using BISM to instrument for general runtime verification cases.
In the second experiment, we instrument a simplified financial transaction system to check for various properties from \cite{BartocciFBCDHJK19}.
In the third experiment, we instrument seven applications from the DaCapo benchmark~\cite{DaCapo06} to verify the classical \textbf{HasNext}, \textbf{UnsafeIterator} and \textbf{SafeSyncMap} properties. 
We compare the performance of BISM, DiSL, and AspectJ in build-time and load-time instrumentation, using three metrics: size, memory footprint, and execution time. 
In build-time instrumentation, the results show that the instrumented code produced by BISM is smaller, incurs less overhead, and its execution incurs less memory footprint.
In load-time instrumentation, the load-time weaving and the execution of the instrumented code are faster with BISM.

This paper extends a paper that appeared in the proceedings of the 20$^{\rm th}$ Runtime Verification conference~\cite{SoueidiKF20}.
The paper provides the following additional contributions:
\begin{itemize}
 \item new language features: meta \locators{}, support for synthetic local arrays, and config files (in \secref{sec:bism-language});
 \item instrumentation model, underlying shadows and equivalence (in \secref{sec:shadows} and \secref{sec:equiv-shadows}).
 \item transformer composition, collision detection and visibility control of advice (in \secref{sec:transformer-composition});
 \item use cases to demonstrate the usage of BISM in different contexts (in Section~\ref{sec:usecases});
 \item an additional experiment demonstrating BISM effectiveness using a benchmark from the competitions on runtime verification~\cite{BartocciFBCDHJK19,RegerHF16} (in \secref{sec:transactions}).
\end{itemize}

\paragraph{Paper organization.}
Section~\ref{sec:bism-design} overviews the design goals and the features of BISM. 
Section~\ref{sec:bism-language} introduces the language of BISM.
 Section~\ref{sec:shadows}, details the instrumentation model of BISM.
Section~\ref{sec:transformer-composition} shows transformers composition in BISM.
Section~\ref{sec:usecases} presents some use cases of BISM in various contexts.
Section~\ref{sec:framework} presents the implementation of BISM. 
Section~\ref{sec:casestudies} reports on the case studies and a comparison between BISM, DiSL, and AspectJ. 
Section~\ref{sec:related-work} discusses related work. 
Section~\ref{sec:conclusion} draws conclusions.
\section{Design Goals and Features}
\label{sec:bism-design}
%
%
BISM is a bytecode instrumentation tool for Java programs implemented on top of ASM~\cite{BrunetonASM02}.
It provides an instrumentation language that is expressive like ASM and provides the user with a high level of abstraction as inspired by AOP. 
BISM is a tool on which RV tools can rely to perform efficient and expressive instrumentation.
In this section, we describe the design goals and features of BISM.
\paragraph{Simple instrumentation model.}
BISM provides an instrumentation model that is easy to understand and use (see~\secref{sec:shadows}).
In particular, BISM does not have the whole notion of pointcuts that allows specifying a predicate to match different joinpoints.
Instead, BISM offers a fixed set of \emph{\locators{}} that capture granular joinpoints.
Each \locator{} is associated with a well-defined region in the bytecode, such as a single instruction, basic block, control-flow branch, or method.
We believe that the \locators{} in BISM can capture the most used joinpoints in the runtime verification of Java programs. 
\paragraph{Instrumentation mechanism.} 
BISM provides a mechanism to write separate instrumentation classes in standard Java. 
An instrumentation class in BISM, which we refer to as a \textit{transformer}, encapsulates the instrumentation logic that is the joinpoint selection and the advice to be injected into the base program.
Advice is specified using special advice instrumentation methods provided by the BISM language that allows bytecode insertion, method invocation, and printing.
\paragraph{Access to program context.}
BISM offers rich access to complete static information about instructions, basic blocks, methods, and classes. 
It also offers dynamic context objects that provide access to values that will only be available at runtime, such as local variables, stack values, method arguments, and results. 
Moreover, BISM allows accessing instance and static fields of these objects. 
Furthermore, new local variables and arrays can be created within the scope of a method to pass values needed for instrumentation.
\paragraph{Control-flow context.}
BISM generates the CFGs of target methods out-of-the-box and offers this information to the user. 
In addition to basic block entry and exit \locators{}, BISM provides specific control-flow related \locators{} to capture conditional jump branches. 
Moreover, it provides a variety of control-flow properties within the static context objects. 
For example, it is possible to traverse the CFG of a method to retrieve the successors and the predecessors of basic blocks.
Edges in CFGs are labeled to distinguish between the True and False branches of a conditional jump.
Furthermore, BISM provides an option to display the CFGs of methods before and after instrumentation, which provides developers with visual assistance for analysis and insights on how to instrument the code and optimize it.
\paragraph{Compatibility with ASM.}
BISM uses ASM extensively and relays all its generated class representations within the static context objects. 
Furthermore, it allows for inserting raw bytecode instructions by using the ASM data types. 
In this case, it is the responsibility of the user to write instrumentation code free from errors. 
If the user unintentionally inserts faulty instructions, the instrumentation may fail. 
The ability to insert ASM instructions provides highly expressive instrumentation capabilities, especially when it comes to inlining the monitor code into the base program, but comes with the cost of possibly producing unwanted behavior.
\paragraph{Instrumentation modes.}
BISM can run in two modes: \emph{build-time} and \emph{load-time}.
In build-time, BISM acts as a standalone application capable of instrumenting all the compiled classes and methods\footnote{Excluding the native and abstract methods, as they do not have bytecode representation.}. 
In load-time, BISM acts as an agent (using JVM instrumentation capability\footnote{The \inlineJava{java.lang.instrument} package.}) that intercepts all classes loaded by the JVM and instruments before the linking phase. 
The load-time mode permits to:
\begin{itemize}
 \item instrument additional classes, including classes from the Java class library that are flagged as modifiable\footnote{The modifiable flag keeps certain core classes outside the scope of BISM and of instrumentation. To the best of our knowledge, there is no exhaustive list of classes with the before-mentioned flag.};
 \item easily performs dynamic program analysis (e.g., profiling, debugging).
\end{itemize}
Instrumentation modes are complementary.
BISM produces a new statically instrumented standalone program in build-time mode, whereas, in the load-time mode, BISM acts as an interface between the program and the JVM (keeping the base program unmodified).
It is generally faster to execute the instrumented program than to load BISM as an agent. 
\paragraph{Portability and ease of use.}
BISM is a lightweight tool designed in Java and fitting in a single jar application of less than 1Mo.
It is hardware-agnostic and only relies on the presence of a JVM in the host software.
The user only needs to include the BISM jar file path to the \emph{classpath} (Java runtime environment variables) to compile new custom transformers.
BISM has been successfully tested on various operating systems and even embedded devices.
As it has no installation requirements, we believe it is portable and usable in any environment.  
\section{BISM Language}
\label{sec:bism-language}
In this section, we present the language of BISM.
The language allows the user to select \emph{joinpoints} (points in the program execution), retrieve relevant context information, and inject advice (i.e., extra code) that can extract information from these points or alter the behavior of the program.

Instrumentation in BISM is specified in Java classes named \textit{transformers}.
BISM language provides \emph{\locators{}} (\secref{sec:locators}) to select joinpoints of interest, static and dynamic context objects (\secref{sec:staticcontext} and~\secref{sec:dynamiccontext}) which retrieve relevant information from these points, and \emph{advice} methods (\secref{sec:instrumentationmethods}) to specify the code to be injected into the base program. 
%
\subsection{\Locators}
\label{sec:locators}
%
\Locators{} provide a mechanism to select joinpoints and specify the advice.
They are implementable methods where the user writes the instrumentation logic.
BISM provides a fixed set of \locators{} classified into four categories: instruction, basic block, method, and meta \locators{}.
%
%
We list below the set of available \locators{} and specify the execution they capture. 
 \paragraph{Instruction.} BISM provides instruction-related \locators{}: 
 \begin{itemize}
 \item \inlineJava{BeforeInstruction} captures execution before a bytecode instruction. 
 \item \inlineJava{AfterInstruction} captures execution after a bytecode instruction. If the instruction is the exit point of a basic block, it captures the code before the instruction. 
 \item \inlineJava{BeforeMethodCall} captures execution just before a method call instruction and after loading all needed values on the stack. 
 \item \inlineJava{AfterMethodCall} captures execution immediately after a method call instruction and before storing the return value from the stack, if any. 
 \end{itemize}
\paragraph{Method.} BISM also provides two method-related \locators{}: 
 \begin{itemize}
 \item \inlineJava{OnMethodEnter} captures execution on method entry block, same execution rules as OnBasicBlockEnter. 
 \item \inlineJava{OnMethodExit} captures execution on all exit blocks of a method before the return instruction.
 \end{itemize}
\paragraph{Basic block.} In addition to the previous \locators{}, BISM provides basic block-related \locators{} which ease capturing control-flow related execution points: 
\begin{itemize}
 \item \inlineJava{OnBasicBlockEnter} captures execution when entering the block, before the first real instruction\footnote{Real instructions are instructions that actually get executed, as opposed to some special Java bytecode instructions such as labels or line number instructions.}.
 \item 
 \inlineJava{OnBasicBlockExit} captures execution after the last instruction of a basic block; except when last instruction is a JUMP/RETURN/THROW instruction, then it captures the code before the instruction. 
 \item 
 \inlineJava{OnTrueBranchEnter} captures execution on the entry of a successor block after a conditional jump on True evaluation. 
 \item 
 \inlineJava{OnFalseBranchEnter} captures execution on the entry of a successor block after a conditional jump on False evaluation. 
\end{itemize}
\paragraph{Meta \locators.} Finally, BISM provides two class related meta-\locators{}: \inlineJava{OnClassEnter} and \inlineJava{OnClassExit}. 
These \locators{} do not capture execution points but can be used for introductions, such as adding new members to a class. 
%
They can also be used to optionally initialize and finalize the transformer execution for each instrumented class.

\begin{figure}
  \centering
 \includegraphics[width=0.45\textwidth]{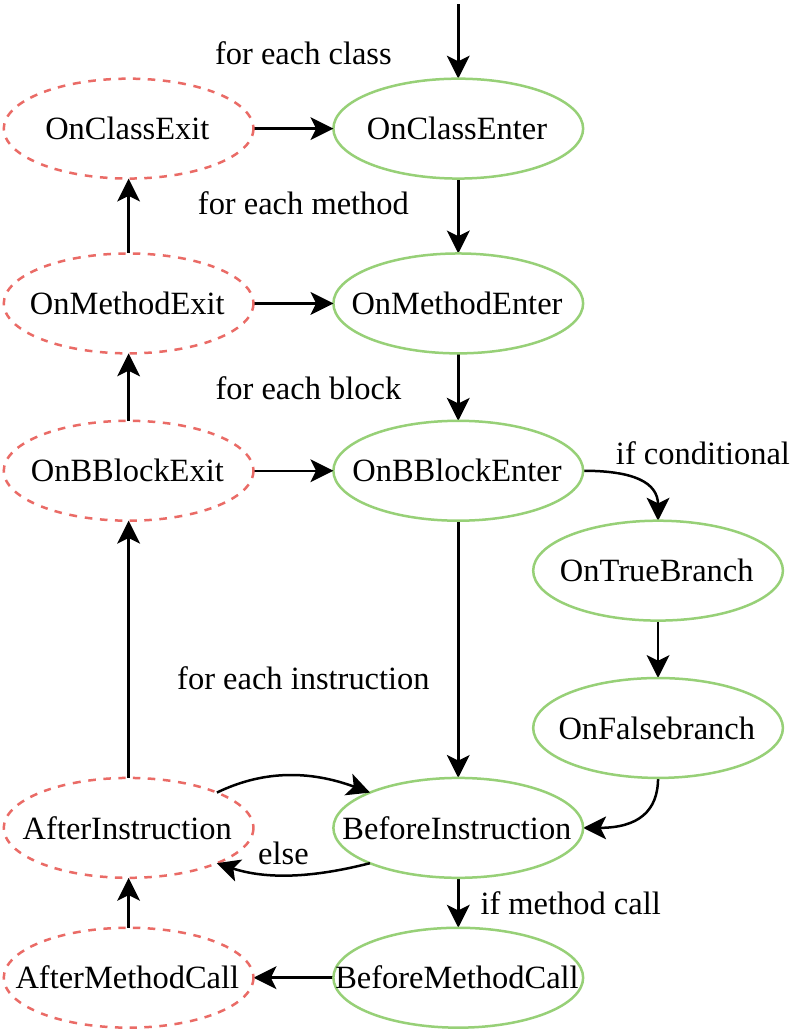}
 \caption{Instrumentation Loop of BISM.}
 \label{fig:instr-loop}
\end{figure}

The order at which \locators{} are visited when applying a transformer is depicted in Figure~\ref{fig:instr-loop}.
Knowing this traversal flow helps the developer know in which order the advice weaving happens.

%
\subsection{Static Context}
\label{sec:staticcontext}
%
\begin{figure}[t]
\centering
 \includegraphics[width=0.65\linewidth]{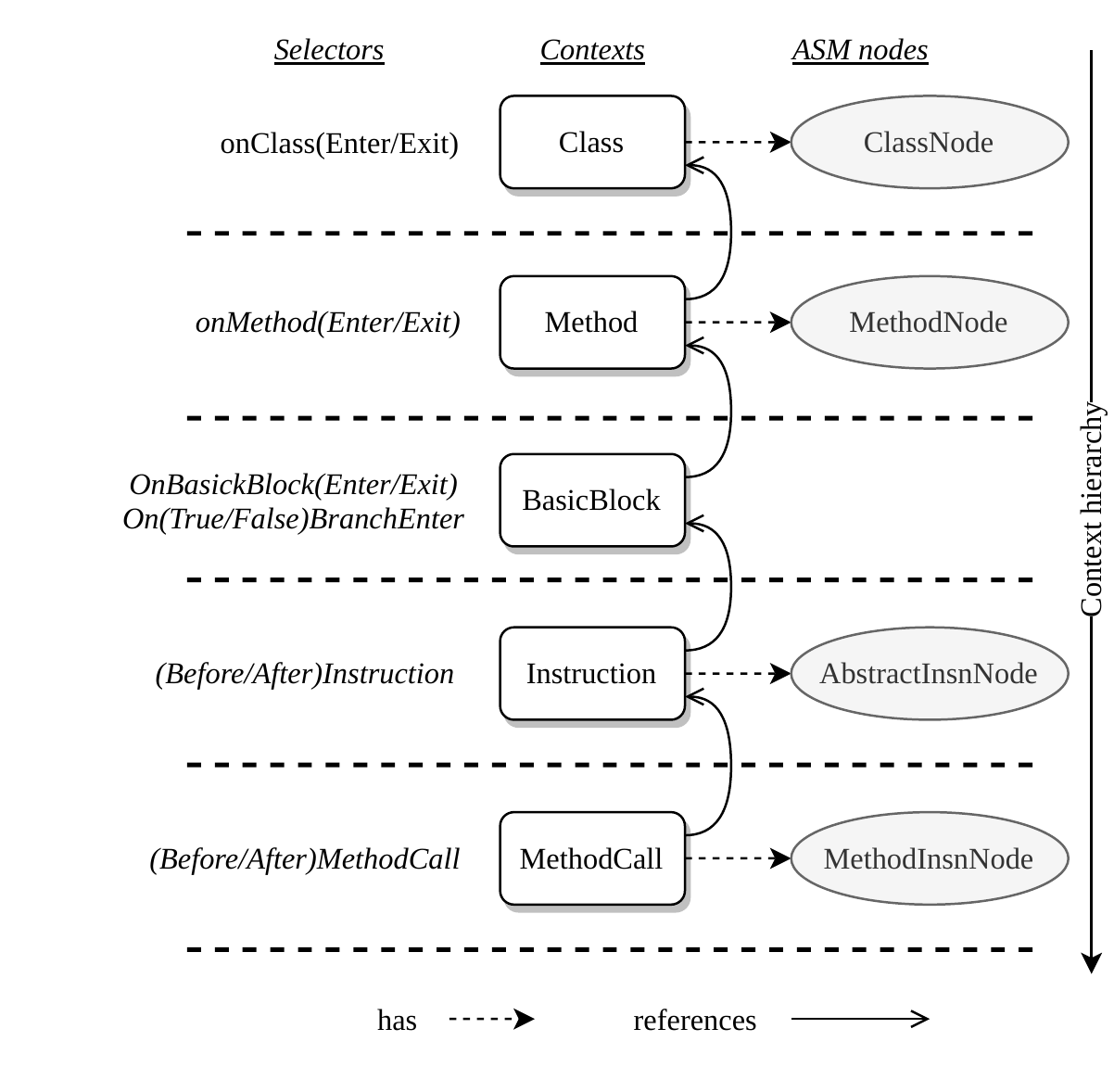}
 \caption{The static context tree related to \locators{} and ASM nodes.}
 \label{fig:context-tree}
\end{figure}
Static-context objects provide access to relevant static information for captured joinpoints in \locators{}. 
Each \locator{} has a specific static context object based on its category.
These objects can be used to retrieve information about bytecode instructions, method calls, basic blocks, methods, and classes.
BISM performs static analysis on the base program and provides additional control-flow-related static information such as basic block successors and predecessors. 
The rich set of context information allows the user to have an expressive joinpoint selection mechanism from within \locators{}.
Unlike AspectJ, BISM does not offer regular expressions to select joinpoints, but from the context objects, one can retrieve the method signature and therefore make the selection manually.
It is even possible to be more selective as BISM offers directly the static context of each bytecode instruction, which is not accessible in AspectJ.
The static context hierarchy and accessibility are summarized in Figure~\ref{fig:context-tree}. 
Each context provides access to the corresponding ASM node object and is accessible from the corresponding \locator{} or by traversing the hierarchy bottom-up as demonstrated in Listing \ref{lst:context-tree}.

\begin{lstlisting}[style=UseCase, label=lst:context-tree, caption=Static context objects hierarchy.]
public void beforeMethodCall(MethodCall mc, ...){
    Instruction i = mc.ins; //Access the surrounding Instruction static context
    BasicBlock b = i.basicBlock; //Access the surrounding BasicBlock context
    Method m = b.method; //Access the surrounding Method context
    ClassContext c = m.classContext; //Access the surrounding Class context
}
\end{lstlisting}
In the following, we detail commonly used properties for each context.
\paragraph*{Common properties.}
At each \locator{} we need to identify the currently instrumented object and its location.
Static context objects contain some identifiers:
\begin{itemize}
 \item a reference to its parent context, named after the parent type;
 \item multiple string identifiers for class and methods, such as name and signature;
 \item a unique (method-wise) integer identifier for basic blocks and instructions.
\end{itemize}

\paragraph{Instruction context.} The \inlineJava{Instruction} context provides relevant information about a single instruction:

\begin{itemize}
 \item \inlineJava{opcode}: its opcode in the JVM instruction set;
 \item \inlineJava{next/previous}: neighbor instructions in the current basic block if they exist;
 \item \inlineJava{isBranchingInstruction()}: indicator of whether it is a branching instruction (multiple successors). BISM takes care of comparing with the right opcodes and ASM node types.
\end{itemize}

It is also possible to retrieve information about the stack context of an instruction, as the information is embedded into the class file :

\begin{itemize}
 \item \inlineJava{getBasicValueFrame()}: returns a list of all local variables, stack items, and their types at the stack frame before executing the current instruction;
 \item \inlineJava{getSourceValueFrame()}: returns a list of all local variables and stack items and their source i.e. which instruction created/manipulated them.
 \end{itemize}


 
%
\paragraph{MethodCall context.} This is a special type of \inlineJava{Instruction} context (only available in \inlineJava{*MethodCall} \locators{}). 
In addition to its \inlineJava{Instruction} context, some specific information is provided, such as the caller method name or the called method class and name. 
%

 %
\paragraph{BasicBlock context.} The \inlineJava{BasicBlock} context provides information about a basic block and its neighborhood in the CFG:
\begin{itemize}
 \item \inlineJava{blockType}: a type to easily identify its role (Entry, exit, conditional, or normal block);
 \item \inlineJava{getSuccessor/PredecessorBlocks()}: all successors and predecessors of the basic block as per the CFG;
 \item \inlineJava{getTrue/FalseBranch()}: the target block after this conditional block evaluates to true (false);
 \item \inlineJava{getFirst/LastRealInstruction()}: the first and last instructions of this block which are executable ones (not labels);
\end{itemize}

\paragraph{Method context.} The \inlineJava{Method} context provides information about the currently visited method:
\begin{itemize}
 \item \inlineJava{name}: the name of the method (not fully qualified);
 \item \inlineJava{getEntryBlock/getExitBlocks()}: the first and last blocks of the method;
 \item \inlineJava{isAnnotated(String)}: checks for an annotation on the method;
 \item some signature information about the method such as its return type and list of formal arguments.
\end{itemize}

\paragraph{Class context.}
The \inlineJava{Class} context provides the name and ASM node of the currently instrumented class.
 
\begin{lstlisting}[style=UseCase,
 caption={A transformer for intercepting basic block executions.},
 label=lst:staticcontextexample 
 ]
public class BasicBlockTransformer extends Transformer {
  @Override
  public void onBasicBlockEnter(BasicBlock bb){
    String blockId = bb.method.className+"."+ bb.method.name+"."+bb.id;
    print("Entered block:" + blockId)
  }

  @Override
  public void onBasicBlockExit(BasicBlock bb){
    String blockId = bb.method.className+"."+ bb.method.name+"."+bb.id;
    print("Exited block:" + blockId)
  }
}
\end{lstlisting}
In Listing~\ref{lst:staticcontextexample}, the transformer uses two \locators{} to intercept all basic block executions (\inlineJava{onBasicBlockEnter} and \inlineJava{onBasicBlockExit}). 
\inlineJava{BasicBlock bb} is used to get the block id, the method name, and the class name. 
The advice method \inlineJava{print} inserts a print invocation in the base program before and after every basic block execution. 
%
\subsection{Dynamic Contexts}
\label{sec:dynamiccontext}
%
BISM also provides dynamic context objects at \locators{} to extract joinpoint dynamic information.
These objects can access dynamic values from captured joinpoints that are possibly only known during the base program execution.
BISM gathers this information from local variables and operand stack, then weaves the necessary code to extract this information. 
In some cases (e.g., when accessing stack values), BISM might instrument additional local variables to store them for later use.
We report here some useful methods, but more are available in the online documentation~\cite{bism}.
For brevity we omit the return type of the methods which is always a \inlineJava{DynamicValue}:

\begin{itemize}
 \item \inlineJava{getThis()}: returns a reference to the class owner of the method being instrumented, and null if the class or method is static;
 \item \inlineJava{getThreadName()}: returns a reference to the name of the thread executing the method being instrumented;
 \item \inlineJava{getLocalVariable(int)}: returns a reference to a local variable by index;
 \item \inlineJava{getStackValue(int)}: returns a reference to a value on the stack by index;
 \item \inlineJava{getStatic/InstanceField(String)}: returns a reference to an instance/static field in the class being instrumented. 
\end{itemize}
BISM gives access to method-relative information.
The runtime arguments passed to a method can be retrieved using \inlineJava{getMethodArgs(int)}.
The method result (return value) can be retrieved using \inlineJava{getMethodResult()}.
The object on which the method is called can be retrieved using \inlineJava{getMethodReceiver()}.

It is also possible to add new local variables of primitive types with the call \inlineJava{addLocalVariable(Object value)}
The scope of the added variables is the method where they are created. 
This is useful for different purposes like to pass data across \locators{}.
Local arrays could also be added with the \inlineJava{createLocalArray(Method, Class)} method.
BISM weaves the necessary bytecode and returns a dynamic value to query and update freely, such as clearing and appending elements.
It is particularly useful when there is a need to pass objects between \locators{} in a method without knowing how much space will be needed at runtime.

Listing~\ref{lst:dynamiccontextexample} presents a transformer the uses the \locator{} \inlineJava{afterMethodCall} to capture the return of an \inlineJava{Iterator} created from a \inlineJava{List} object.
It uses the dynamic context object using \inlineJava{MethodCallDynamicContext dc} provided to the \locator{} to retrieve the dynamic data.
The example also shows how to limit the scope to a specific method using an if-statement on the static context.

\begin{lstlisting}[style=UseCase,
 caption={A transformer that intercepts the creation of an iterator from a \inlineJava{List}.},
 label=lst:dynamiccontextexample
 ]
@Override
public void afterMethodCall(MethodCall mc, MethodCallDynamicContext dc){
  if (mc.methodName.equals("iterator") && mc.methodOwner.endsWith("List")) {
      //Access to dynamic data
      DynamicValue callingClass = dc.getThis(mc);
      DynamicValue list = dc.getMethodTarget(mc);
      DynamicValue iterator = dc.getMethodResult(mc);

      //Invoking a monitor
      StaticInvocation sti = new StaticInvocation("IteratorMonitor", "iteratorCreation");
      sti.addParameter(callingClass);
      sti.addParameter(list);
      sti.addParameter(iterator);
      invoke(sti);
  }
}
\end{lstlisting}

\subsection{Advice Methods}
\label{sec:instrumentationmethods}

A user inserts advice into the base program at the captured joinpoints using the advice instrumentation methods.
Advice methods allow the user to extract needed static and dynamic information from within joinpoints, also allowing arbitrary bytecode insertion.
These methods are invoked within \locators{}.
BISM provides \inlineJava{print} methods with multiple options to invoke a print command.
It also provides (i) \inlineJava{invoke} methods for static method invocation and (ii) \inlineJava{insert} methods for inserting bytecode instructions.
These methods are compiled by BISM into bytecode instructions and inlined at the referenced bytecode location.
%
We list below the advice methods available in BISM. 

\paragraph{Printing on the console.}
Instrumenting print statements in the base program can be achieved via method \inlineJava{print}, which permits to write both on the standard and error output of the base program.
%
These methods take either static values or dynamic values retrieved in \locators{}.
Listing \ref{lst:staticcontextexample} shows an example of using one of the print helper methods to instrument the base program to print the basic block constructed id.

\paragraph{Invoking static methods.} 
Invoking external static methods can be achieved using the advice method \inlineJava{invoke}. 
An object of type \inlineJava{StaticInvocation} should be constructed and provided with the external class name, the method name, and parameters.
Listing \ref{lst:dynamiccontextexample} depicts a transformer that instrument the base program to call an external static method (here named \inlineJava{iteratorCreation}).
The \inlineJava{StaticInvocation} constructor takes the class and method names as input.
Parameters can be added using \inlineJava{addParameter()}.
It supports either \inlineJava{DynamicValue} type or any primitive type in Java, including \inlineJava{String} type (any other type will be ignored). 
After that, \inlineJava{invoke} weaves the method call in the base program.

\paragraph{Raw bytecode instructions.} 
Inserting raw bytecode instructions can be achieved with the \inlineJava{insert} methods.
When used, the developer's responsibility is to write correct instructions that respect the JVM static and structural constraints. 
Errors can be introduced by ignoring the stack requirements and altering local variables.
For Java 8 and above programs, using the insert methods to push new values on the stack or create local variables requires modifying
the \emph{maxStack} and \emph{maxLocals} values.
All static contexts give access to the needed ASM object \inlineJava{MethodNode} to increment the values maxLocals and maxStack from within the joinpoint.

\subsection{Instrumentation Scoping}
\label{sec:runtime-args}

BISM provides many configuration features, such as limiting the scope of the instrumentation or passing arguments to the transformers to modify their behavior.
For example, the \emph{scope} global argument permits matching classes and methods by their names.

Specifying \emph{(scope=java.util.List.*, java.util.Iterator.next)} will instrument all methods in the \inlineJava{List} class and only the next method in the \inlineJava{Iterator} class.
Moreover, static context objects can also be used to limit the scope of instrumentation from inside \locators{}; they can provide more precise scoping information demonstrated in Listing \ref{lst:dynamiccontextexample}.
It is recommended using the \emph{scope} argument when possible to avoid analyzing unwanted classes, enhancing instrumentation performance.

\subsection{User Configuration}
\label{sec:user-config}
To favor usability, BISM execution accepts arguments both from the command line (which has higher priority) or through a configuration file.
Configurable settings such as printing the CFG files and dumping the instrumented bytecode can be specified.
The configuration file is more expressive as it also permits passing arguments to transformers.
A transformer may need arguments to modify its internal behavior, e.g., a flag for logging.

\section{Instrumentation Model}
\label{sec:shadows}
%
In this section, we introduce the instrumentation model implemented in BISM.
We show the correspondence between joinpoints captured by \locators{} and bytecode regions in the program given by \emph{shadows}.
We then introduce the equivalence between shadow, which is used when composing transformers (\secref{sec:transformer-composition}).
%
\subsection{Overview}
%
A \emph{joinpoint} is essentially a configuration of the base program traversed during its execution.
A joinpoint consists of static and dynamic context information.
The static context of a joinpoint is defined by a lexical part in the source code.
The dynamic context is made of runtime information available in the stack and memory.
Depending on the instrumentation task, the user is only interested in some of the joinpoints generated by the program execution. 
At these joinpoints, the user-specified advice is executed.

When instrumenting a program, we inject extra instructions at specific bytecode locations in the code of the base program.
Bytecode regions denote the locations where instrumentation can happen.
The user implements \locators{} (\secref{sec:locators}) to select joinpoints using lexical conditions that mark bytecode regions.
Lexical conditions are given by \emph{shadows}.
A shadow is a pair defined by a lexical element (instruction, method, basic block) identifier and a direction.
Shadows are bytecode instructions along with the direction (\textit{before} and \textit{after}), or basic blocks and methods along with the \textit{enter} or \textit{exit} direction. 
A shadow refers to one bytecode region, and a bytecode region can be referred to by multiple shadows.
The joinpoints selection in BISM is statically determinable.
That is, determining whether a joinpoint is selected is statically decidable.
%
\subsection{Shadows}
%
Shadows are constructs used to mark the bytecode regions in the base program where instrumentation can happen.
BISM internally operates on shadows to extract static information and delimit the regions where advice will be woven.
Shadows are pairs consisting of bytecode instructions along with a specified direction (\textit{before} and \textit{after}), or basic blocks and methods along with an \textit{enter} or \textit{exit} direction. 

Let $\Methods$ be the set of all methods that have bytecode representation, $\Blocks$ be the set of all basic blocks, and $\Instrs$ be the set of all bytecode instructions in the base program. 
Then, $\Shadows$ represents the set of all shadows in a program that BISM identifies in a program.
\begin{align*}
 \Shadows = & \quad \ (\{\mathrm{before}, \mathrm{after}\} \times \Instrs) \\
 & \cup \  (\{ \mathrm{enter}, \mathrm{exit}\} \times \Blocks) \\
 & \cup \ (\{ \mathrm{enter}, \mathrm{exit}\} \times \Methods)
\end{align*}


For a method $m \in \mathit{Methods}$, $m.\blocks$ $\subseteq$ $\Blocks$ denote the basic blocks in the CFG of $m$, and $m.\instrs \subseteq \Instrs$ denote the indexed list of instructions in $m$. 
\begin{definition}[Method Shadows]
 \label{def:methodshadows}
 $\Shadows_m$ denotes the set of all shadows in a method.
 \begin{align*}
 \Shadows_m = & \quad \ (\{\mathrm{before}, \mathrm{after}\} \times m.\instrs) \\
 & \cup \  (\{ \mathrm{enter}, \mathrm{exit}\} \times m.\blocks) \\
 & \cup \ \{ \tuple{\mathrm{enter}, m}, \tuple{\mathrm{exit},m}\} 
 \end{align*}
 
 \end{definition}

The shadows of a method are restricted to its instructions and basic blocks. 
We give an example of the shadows in a method as identified by BISM. 

\begin{lstlisting}[style=UseCase, label=lst:methodm, caption=A method calling an Iterator.]
public void m() {
    //Initialize a list of strings
    List<String> l = new ArrayList<>();
    l.add("A");
    //Create iterator
    Iterator<String> i = l.iterator();
    //Call next if iterator has next
    if (i.hasNext())
        i.next();

    System.out.print("done");
} 
 \end{lstlisting}

 \begin{lstlisting}[float=H!, style=UseCase, label=lst:bytecode, caption=Bytecode and associated shadows for the method in Listing~\ref{lst:methodm}., mathescape=true] 
 m() {
 $\green{\tuple{\mathrm{enter},m_0}}$ 
 $\blue{\tuple{\mathrm{enter},b_0}}$ 
 $\green{\tuple{\mathrm{before},i_0}}$ 
 _new ArrayList 
 $\green{\tuple{\mathrm{after},i_0}}$ 
 $\green{\tuple{\mathrm{before},i_1}}$ 
 dup 
 invokespecial ArrayList.init ()V
 astore 1 
 aload 1 
 ldc A 
 invoke List.add (Object;)Z 
 pop 
 aload 1 
 $\red{\tuple{\mathrm{before},i_9}}$ 
 invokeinterface List.iterator ()Iterator; 
 $\green{\tuple{\mathrm{after},i_9}}$ 
 astore 2 
 aload 2 
 invokeinterface Iterator.hasNext ()Z 
 $\blue{\tuple{\mathrm{exit},b_0}}$ 
 ifeq L0
 $\blue{\tuple{\mathrm{enter},b_1}}$ 
 aload 2 
 invokeinterface Iterator.next ()Object; 
 pop 
 $\blue{\tuple{\mathrm{exit},b_1}}$
 L0
 $\blue{\tuple{\mathrm{enter},b_2}}$
 getstatic System.out, PrintStream; 
 ldc done 
 invokevirtual PrintStream.print (String;)V
 $\blue{\tuple{\mathrm{exit},b_2}}$ 
 $\green{\tuple{\mathrm{exit},m_0}}$
 return
 }
 \end{lstlisting}

\begin{example}[Method shadows]
\label{ex:shadows}
Listing~\ref{lst:methodm} contains a Java me\-thod \inlineJava{m} that creates a \inlineJava{List l} with an associated \inlineJava{Iterator i}.
The method checks if \inlineJava{i.hasNext()} and calls \inlineJava{i.next()}.
Listing~\ref{lst:bytecode} shows the (simplified) bytecode for method $m$ in black font.
We also show the shadows identified by BISM, added in the colors blue, red, and olive green.
There are two shadows for the method entry, and exit points showed.
Two shadows for each basic block (the if-statement results in having three basic blocks), and for each instruction, two shadows to delimit the region before it and after it (we omitted most of the instruction shadows for brevity).
\end{example}
%
\subsection{\Locators{} Matching Shadows}
\label{sec:locator-shadows}
%
\sloppy
Each BISM \locator{} matches a specific subset of the shadows.
\Locators{} \inlineJava{OnMethodEnter} and \inlineJava{OnMethodExit} respectively match the shadows $\tuple{\mathrm{enter},m}$ and $\tuple{\mathrm{exit},m}$ for each method $m \in \Methods$.
\Locators{} \inlineJava{OnBasicBlockEnter} and \inlineJava{OnBasicBlockExit} respectively match the shadows $\tuple{\mathrm{enter},b}$ and $\tuple{\mathrm{exit},b}$ for each basic block $b \in \Blocks$.
\Locators{} \inlineJava{OnTrueBranch} and \inlineJava{OnFalseBranch} match special instances of the shadows $\tuple{\mathrm{enter},b}$ where basic block $b$ has a predecessor block ending with a conditional jump.
\Locators{} \inlineJava{BeforeInstruction} and \inlineJava{AfterInstruction} respectively match the shadows $\tuple{\mathrm{before},i}$ and $\tuple{\mathrm{after},i}$ for each instruction $i \in \Instrs$.
\Locators{} \inlineJava{BeforeMethodCall} and \inlineJava{AfterMethodCall} respectively match the shadows $\tuple{\mathrm{before},i}$ and $\tuple{\mathrm{after},i}$ for each instruction $i \in \Instrs$ that is a method invocation instruction.
We demonstrate the matching in the following example.
\begin{example}[\Locators{} matching shadows]
 \label{ex:shadows-locators}
 In this example, we show the shadows matched by BISM \locators{} when applying different transformers to the Java program from Example~\ref{def:methodshadows}.
 Applying the transformer from Listing~\ref{lst:staticcontextexample}, the \locators{} \inlineJava{OnBasicBlockEnter} and \inlineJava{OnBasicBlockExit} match the shadows highlighted in blue in Listing~\ref{lst:bytecode}.
 The method has three basic blocks because of the if-statement.
 These shadows mark the regions where BISM weaves the code to print the basic block id.
 Applying the transformer from Listing~\ref{lst:dynamiccontextexample}, the \locator \inlineJava{afterMethodCall} matches the shadow highlighted in the color red in Listing~\ref{lst:bytecode}.
 This shadow marks the region where BISM weaves the code to call the monitor \inlineJava{IteratorMonitor.iteratorCreation}.
\end{example}
\begin{remark}[Shadows, traversal, and static analysis]
 One of the advantages of the shadows identified by BISM is that they can be used as a traversal strategy for the base program.
 The traversal strategy can be seen in Figure~\ref{fig:instr-loop}.
 This allows the user to implement compile-time static analyzers, as we will see in~\secref{sec:usecases}.
 A transformer can be implemented to analyze the code without instrumenting it.
 \Locators{} play the role of Visitor methods where the user can write custom code for the static analyzer without even instrumenting the base program using the advice methods(\secref{sec:instrumentationmethods}).
 This enables combining static analysis with runtime verification using transformer composition (\secref{sec:transformer-composition}) and helps in optimizing instrumentation.
\end{remark}
%
\subsection{Equivalence Between Shadows}
\label{sec:equiv-shadows}
%
Since a bytecode region can be referred to by multiple shadows, to detect when two transformers contain \locators{} with shadows that mark the same regions, we define the notion of equivalence between shadows.
%
The user may unintentionally target the same regions using different \locators{} in a transformer.
The choice of \locator{} used depends on the static/dynamic context needed by the user; since each \locator{} exposes different static/dynamic context objects.
We define the equivalence relation over shadows below and give an illustrative example.

\begin{definition}[Equivalence Relation over Shadows]
 \label{def:equivrelation}
 The equivalence relation over shadows in a method $m$ is defined as follows: 
 \begin{equation*} \label{eq1}
 \begin{split}
 \shadowequiv & \subseteq \Shadows_m \times \Shadows_m 
 \end{split}
\end{equation*} 
 \begin{align*}
 \defas & \ \ \  \  \{ \tuple{\mathrm{enter},m}, \tuple{\mathrm{enter},m.\mathrm{entryBlock}} \ \} \tag{1} \\
 & \cup \{ \ \tuple{\mathrm{exit},b} , \tuple{\mathrm{exit},m} \ | \  b \in m.\mathrm{exitBlocks} \ \} \tag{2} \\
 & \cup \{ \ \tuple{\mathrm{enter},b}, \tuple{\mathrm{before},i} \ | \ b \in m.\blocks \ \wedge   \ i.\mathrm{index} = b.\mathrm{first.index} \} \tag{3}\\
 & \cup \{ \ \tuple{\mathrm{after},i}, \tuple{\mathrm{exit},b} \  | \ b \in m.\mathrm{blocks}  \ \wedge \  i.\mathrm{index} = b.\mathrm{last.index} \} \tag{4} \\
 & \cup \{ \ \tuple{\mathrm{after},i}, \tuple{\mathrm{before},i^\prime} \ | \ \exists k \in [1,\mathrm{size}(m.\mathrm{instrs})]:  i.\mathrm{index} = k \ \wedge \  i^{\prime}.\mathrm{index} = k+1 \} \tag{5}
 \end{align*} 

\end{definition}

In Definition \ref{def:equivrelation}, line (1) states that the region on method enter is equivalent to the region on basic block enter if the block is the entry block of the method.
Line (2) states that the region at a method exit is equivalent to the region at the exits of all basic blocks that are exit blocks in the method.
Line (3) states that the region at block entry is equivalent to the region before the first instruction in the block defined by $b.\mathrm{first}$.
Line (4) states that the region at the exit of a block is equivalent to the region after the last instruction of the block defined by $b.\mathrm{last}$.
Line (5) states that the region after an instruction and before its consecutive are equivalent.

\begin{example}[Equivalent shadows in a method]
 \label{ex:equivalentShadows}
Figure \ref{fig:shadows}, depicts the CFG of a method $m$ with 4 basic blocks ($b_1$, $b_2$, $b_3$, $b_4$) where $b_1$ is the entry block, $b_2$ and $b_4$ are both exit blocks.
In basic block $b_2$, we show two consecutive instructions $i$ and $j$.
In basic block $b_3$, we show instruction $k$ as the first instruction in the block and instruction $l$ as the last instruction.
The filled grey boxes in the figure illustrate the equivalent shadows, numbered as their corresponding line in Definition \ref{def:equivrelation}.
From (1), we have $\tuple{\mathrm{enter},m} \shadowequiv \tuple{\mathrm{enter},b_1}$.
From (2), we have $\tuple{\mathrm{exit},b_4} \shadowequiv \tuple{\mathrm{exit},b_2} \shadowequiv \tuple{\mathrm{exit},m}$.
From (3), we have $\tuple{\mathrm{enter},b_3} \shadowequiv \tuple{\mathrm{before},k}$.
From (4), we have $\tuple{\mathrm{after},l} \shadowequiv \tuple{\mathrm{exit},b}$.
From (5), we have $\tuple{\mathrm{after},i} \shadowequiv \tuple{\mathrm{before},j}$.

\begin{figure}[t!]
 \centering
 \includegraphics[width=0.45\textwidth]{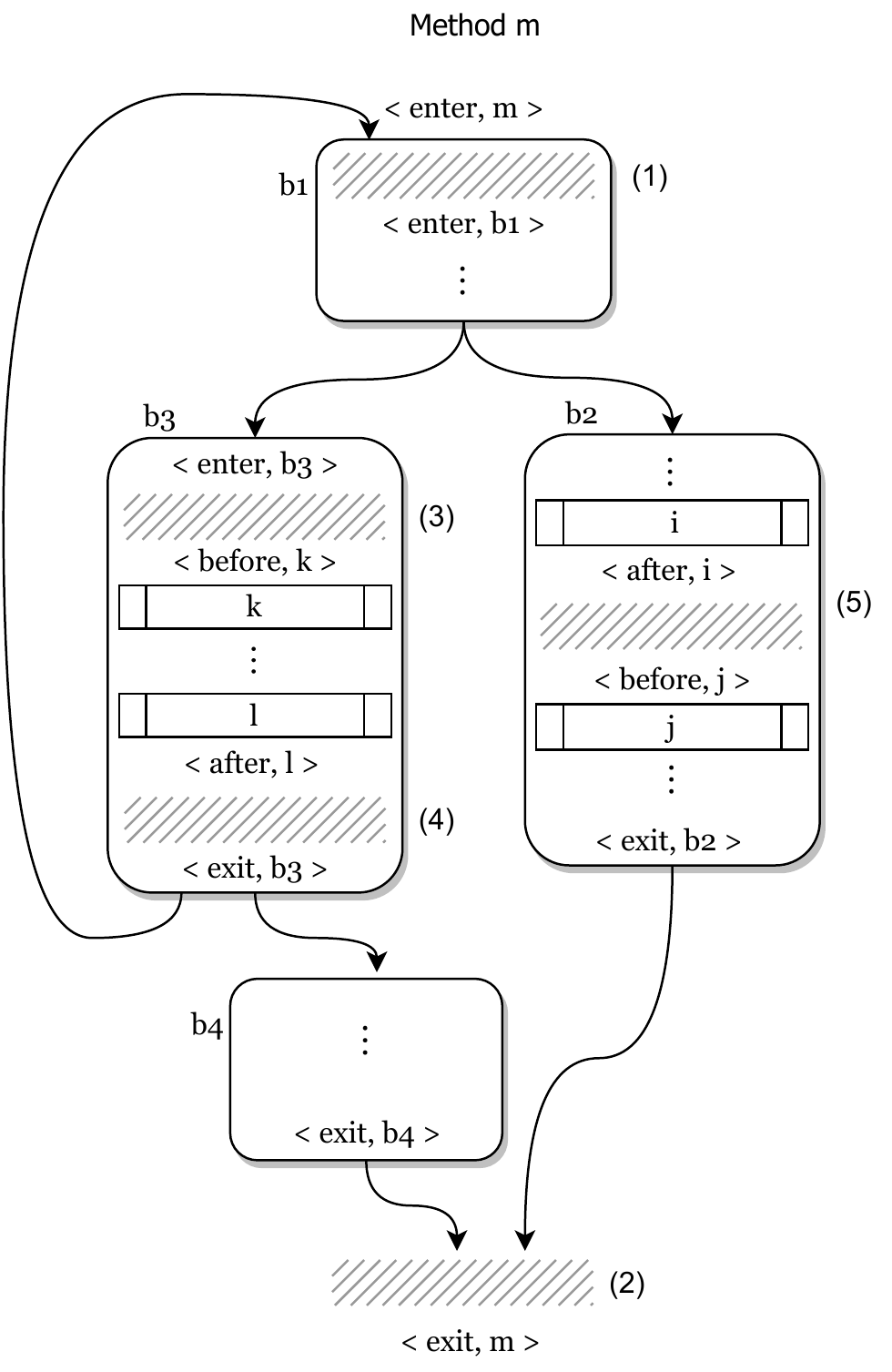} 
 \caption{Illustration of shadows and their equivalence relation.} 
 \label{fig:shadows}
 \end{figure}

\end{example}
\section{Transformer Composition}
\label{sec:transformer-composition}

BISM allows users to apply more than one transformer in a single run.
The transformers are applied sequentially to the base program in the order specified by the user.
We refer to applying multiple transformers in a single run as \emph{transformer composition}.
In this section, we discuss the motivation for composing transformers (\secref{sec:motivation-composition}) and address some concerns that may arise when 
multiple transformers target the exact program bytecode regions (\secref{sec:collision}) and how BISM can help (\secref{sec:order-matters} and \secref{sec:controlling-visbility}).
\subsection{Motivations for Composition}
\label{sec:motivation-composition}

Composition is needed in some cases and optional in others.
Transformer composition is obligatory when it is impossible to merge the code of two transformers in one transformer. 
This situation arises when there is a dependency between transformers, and more than one pass is required to instrument the program.
In other cases, we may want to implement separate transformers based on their functionality for a cleaner code.
We discuss the two cases.
\paragraph{More than one pass.}
In many cases, transformations might require multiple passes on the same class.
Let us assume that we are implementing a simple obfuscator that randomly changes the names of all methods in a program. 
In this case, one pass is not enough; we need one pass to map the original names to the obfuscated names and then another pass to change the classes and method names.
BISM can also be used to implement static analyzers.
This enables plenty of scenarios where static analysis can be leveraged in combination with runtime verification.
In such cases, a transformer can be implemented to perform the analysis and applied before the transformer responsible for instrumenting the code for monitoring.

\paragraph{Modularity of transformers.}
At the core of aspect-oriented programming is increasing modularity by separating concerns.
Hence we encourage separating transformers based on their functionality.
This allows different team members to implement different transformers separately, where a single transformer should logically handle one concern.
Let us say, we want to instrument a program to monitor at runtime safety properties such as \textit{HasNext}, \textit{UnSafeIterator} and \textit{SafeSyncMap} (see \secref{sec:dacapo}).
Implementing a single transformer for each property is more readable and favors reuse.
%
\subsection{Transformer Collision}
\label{sec:collision}
%
Transformers impose new ``aspects" into the base program through inserting advice.
When two transformers insert advice that targets the same program bytecode regions, we say that the two transformers collide.
BISM detects and reports transformer collisions, which makes the composition more transparent to the user.

BISM determines collision detection after weaving the advice of multiple transformers into the base program.
Recall Definition\ref{def:methodshadows} of the shadows of a method.
Let $\Shadows_m^t$ represent all the shadows used by BISM to insert the advice for transformer $t$, in a method $m$, we have: $$\Shadows_m^t \subseteq \Shadows_m$$ 
To detect collision in a method, we check whether two transformers insert advice at equivalent shadows (\secref{sec:equiv-shadows}).

\begin{definition}[Transformer Collision]
 \label{def:collision}
 Transformer $t$ collides with transformer $t^\prime$ in method $m$, iff
 $$ \exists \ s \in \Shadows_m^t \ , \  \exists \ s^\prime \in \Shadows_m^{t^\prime} \ : \  s \shadowequiv s^\prime $$
 
\end{definition}
Notice that the collision between transformers is symmetric, which means that the order of applying two transformers is irrelevant to detect collision.
Also, collision is reflexive, which means that collision is also detected when applying the same transformer twice.
%
 

\paragraph{Collision report.}
{
To detect a collision, we compute equivalent shadows used for instrumentation by transformers.
BISM records the used shadows after weaving each transformer and reports all collisions after each run.
The report shows the exact locations of collision along with the colliding transformers to the user.
}

Several concerns may arise from collisions, such as determining the order of execution and the visibility among aspects.
In the following sections, we discuss these problems.

\subsection{Order Matters}
\label{sec:order-matters}

BISM executes the transformations in the order specified by the user.
One issue that may arise is that applying the transformations in a different order may exhibit different behavior in the final instrumented program. 
Let us look at the following example.

\begin{example}[Order matters]
Listing \ref{lst:transformerComposition}, demonstrates two transformers: \inlineJava{Logging} which is used to log enters and exits to all methods, and \inlineJava{Timer} that profiles the time needed for each method to execute.
We can see that the joinpoints captured by \inlineJava{onMethodEnter} and \inlineJava{onMethodExit} are common between both transformers. Hence the advice will be inserted at the same bytecode regions, and we have a \textit{collision}.
Applying transformer \inlineJava{Logging} before \inlineJava{Timer} results in the timer calculating the original method computation in the base program and the time needed by the logger.
We may want to give precedence to the timer to calculate the time without the logging operations; hence logging will wrap the base program and the timer operations. 

\begin{lstlisting}[style=UseCase, 
 caption={Logging and Timer transformers (written in pseudo-code).},
 label=lst:transformerComposition,
]
public class Logging extends Transformer {
    public void onMethodEnter(..) {
        //Log methodName + Enter
    }
    public void onMethodExit(..) {
        //Log methodName + Exit
    }
}
public class Timer extends Transformer {
    public void onMethodEnter(..) {
        //Initialize a timer object
    }

    public void onMethodExit(..) {
        //Log time elapsed + methodName
    }
}
\end{lstlisting}
\end{example}

In other cases, the execution order is not important to us, even if there is a collision between two transformers. 
In \secref{sec:dacapo}, we implement multiple transformers to instrument and monitor different safety properties.
In this case, changing the order of the transformations does not produce any observable semantic difference in the final instrumented program.
In general, the user is encouraged to check the collision report (\secref{sec:collision}) and manually change the order of transformers if needed.

\subsection{Controlling Visibility}
\label{sec:controlling-visbility}

When composing transformers, each transformer introduces a new set of instructions to the base program.
The newly added instructions are then part of the base program and become visible to the second transformer to use.
In many cases, we may want to hide these instructions in a composition.
BISM provides the attribute \inlineJava{@Hidden} that can be placed as an annotation on a transformer.
When used, the newly added instructions are not intercepted anymore by the \locators{}.
A prevalent scenario is to avoid instrumenting previously added code by a different transformer. Let us look at the following example.
\begin{example}[Hidden transformers]
In Listing \ref{lst:tot}, demonstrates two transformers: \inlineJava{CountMethodCalls} which counts the number of method calls and \inlineJava{LogMethodCalls} which logs all method calls.
We can see that the joinpoints captured by \inlineJava{onMethodCall} are shared between both transformers. Hence the advice will be inserted at the same bytecode regions, and we have a \textit{collision}. 
A user might be only interested in logging the method calls of the base program and does not want the logger to log counting calls introduced by \inlineJava{LogMethodCalls}.
Alternatively, the user might be interested in counting the method calls of the base program and not the log calls introduced by \inlineJava{CountMethodCalls}.
Adding the \inlineJava{@Hidden} attribute on the transformers hides the newly added instructions from other transformers in a composition.

\begin{lstlisting}[style=UseCase,
 caption={\inlineJava{@Hidden} attribute on transformers.},
 label=lst:tot,
]
@Hidden
public class CountMethodCalls extends Transformer {
    public void onMethodCall(..) {
        //Invoke methodCounterIncrement
    }
}
@Hidden
public class LogMethodCalls extends Transformer {
    public void onMethodCall(..) {
        //Invoke logger
    }
}
\end{lstlisting}
\end{example}

\paragraph{Hidden instructions.}

BISM also allows transformers to hide arbitrary instructions of the base program from other transformers by providing a mechanism to mark instructions as hidden. 
When an instruction is marked as hidden, it is excluded from $\Shadows_m$ and thus not exposed to \locators.
Hence, it will not be intercepted by the transformers that follow.
This feature can be used for optimizing instrumentation by having one transformer implementing a static analyzer that hides particular instructions from the instrumentation transformer.

In general, to avoid instrumenting previously added advice, the user is encouraged to check the collision report (\secref{sec:collision}) and use the \inlineJava{@Hidden} when needed.


\section{Some Example Use Cases}
\label{sec:usecases}
%
%
In this section, we show the versatility of BISM by demonstrating some use cases in static and dynamic analysis of programs, namely the mutation, code analysis, runtime verification, dynamic profiling, and logging of programs.

\subsection{Mutation of Programs}
\label{sec:mutation}
%
%
We consider software testing and, more particularly, mutation testing (see~\cite{mutationSurvey} for a survey).
Mutation testing aims to ensure software quality by checking that slightly modified versions of a program (i.e., mutants) will not pass the same tests as the original.
Mutants emulate the programs that would be obtained as the result of programmers' mistakes.
There are various types of mutations of various complexity levels~\cite{liptonMutation,Offutt2001}.
We consider the following types of often occurring mutations:
\begin{itemize}
 \item Value mutations, which change variable values into the program or return values.
 \item Operator mutations, which change the logical or arithmetical operators used across the program.
 \item Statement mutations, which change complex constructions, like method calls or even the CFG of the program.
\end{itemize}
In the following, we define an example mutator for each type of mutation, i.e., a transformer producing such mutations.
%
\subsubsection{Return Mutator: \emph{Value Mutation}}
%
The mutator in Listing~\ref{ret-mut} emulates the fact that a default return value has been forgotten in the program.
Hence, the target method always returns the same fixed value instead of the normally computed one.
For this, the mutator uses the \inlineJava{onMethodExit} joinpoint and detects whether the parameter method \inlineJava{m} returns a value using the method type.
In such a case, the mutator removes the value from the stack.
Then, a fixed value (here 0 for integers) is pushed onto the stack to be returned.

\begin{lstlisting}[style=UseCase, caption={Return mutator.},label=ret-mut]
public void onMethodExit(Method m,...){
//Detecting return type
    if (Type.getReturnType(m.methodNode.desc) == Type.VOID_TYPE)
        return;
    //Remove return value from stack
    if (Type.getReturnType(m.methodNode.desc) .getSize() == 1)
        insert(new InsnNode(Opcodes.POP));
    else
        insert(new InsnNode(Opcodes.POP2));
    //Push fixed return value (0)
    switch(Type.getReturnType(m.methodNode.desc) .getSort()){
        case Type.INT:
            insert(new InsnNode(Opcodes.ICONST_0));
        break;
        ...
    }
}
\end{lstlisting}
%
\subsubsection{Instruction Mutator: \emph{Operator Mutation}}
%
The mutator in Listing~\ref{gen-mut} performs some replacements on a specified set of instructions.
The mutator is generic and relies on some abstract methods.
A replacement instruction can either be randomly chosen or obtained using a user-defined \emph{mapping} between instructions.
To do this for an instruction, the mutators check whether the instruction is in its scope and if so, it replaces it.
%

\begin{lstlisting}[style=UseCase, caption={Generic instruction mutator.}, label=gen-mut]
public void beforeInstruction(Instruction ins, ...){
    if (isCovered(ins)){
        remove(ins);
    if (negate)
        insert(negate(ins));
    else
        insert(random(ins));
    }
}
//Check whether a particular instruction is covered (type, position, ...)
abstract boolean isCovered(Instruction);
//Choose a random operation (compatible in term of type, arg count ...)
abstract AbstractInsnNode random(Instruction);
//Negate the opcode of a given instruction when applicable
abstract AbstractInsnNode negate(Instruction);
\end{lstlisting}

We present two instances of the operator mutator, which are obtained by implementing the abstract methods.
\begin{itemize}
\item 
The mutator in Listing \ref{if-mut} targets \emph{conditional operators}, which are detected as \emph{conditional jump} instructions.
Another comparison operator replaces conditional operators without changing their destination.
\item
The mutator in Listing \ref{int-mut} targets binary arithmetic operators on integers.
Arithmetic operators are replaced either by a random operator or the complementary one ($-$ and $+$, $\&$ and $|$ for bitstring operators\ldots).
\end{itemize}

\begin{lstlisting}[style=UseCase, caption={Decision mutator.}, label=if-mut]
//If it is a conditional
boolean isCovered(Instruction ins) {
    return ins.isConditionalJump();
}

//Choose a random if which is compatible in term of type and arg count
AbstractInsnNode random(Instruction insIf) {
    if (insIf.opcode >= Opcodes.IFEQ && insIf.opcode <= Opcodes.IFLE)
        return new JumpInsnNode(Opcodes.IFEQ + r.nextInt(Opcodes.IFLE - Opcodes.IFEQ+1), ((JumpInsnNode) insIf.node).label);
 ...
}

//Negate the opcode of a given if
AbstractInsnNode negate(Instruction ins){
    if (ins.opcode == Opcodes.IFNULL || ins.opcode % 2 == 1)
        return new JumpInsnNode(ins.opcode +1, ((JumpInsnNode) ins.node).label);
    else
        return new JumpInsnNode(ins.opcode -1, ((JumpInsnNode) ins.node).label);
}
\end{lstlisting}

\begin{lstlisting}[style=UseCase, caption={Arithmetic mutator.}, label=int-mut]
final List<Integer> I2Opcodes = Arrays.asList(
    Opcodes.IADD, Opcodes.ISUB,
    Opcodes.IMUL, Opcodes.IDIV,...);
 
boolean isCovered(Instruction ins){
    //All double int operand arithmetic instructions
    return I2Opcodes.contains(ins.opcode);
}

AbstractInsnNode random(Instruction ins){
    return new InsnNode(I2Opcodes.get((int) (Math.random()*I2Opcodes.size())));
}

AbstractInsnNode negate(Instruction ins){
    return new InsnNode(ins.opcode + ( I2Opcodes.indexOf(ins.opcode) % 2 == 0 ? 1 : -1));
}
\end{lstlisting}

\subsubsection{Void Call Mutator: \emph{Statement Mutation}}

The mutator in Listing~\ref{void-mut} removes calls to methods with the void return type.
For this, whenever there is a call to such a method, the transformer unloads its parameters from the stack and removes the \inlineJava{INVOKEX} opcode.
To check for return types and unload the parameters differently regarding their sizes, the transformer iterates through the method descriptor\footnote{The descriptor is a string representing a type, for a method it permits to access the return and argument types.} available through the static context attribute \inlineJava{mc.methodnode.desc} .

\begin{lstlisting}[style=UseCase, caption={Void call mutator.}, label=void-mut]
public void beforeMethodCall(MethodCall mc,...){
    if (Type.getReturnType(mc.methodnode.desc) != Type.VOID_TYPE)
        return;
    //Pop each argument, respecting its size
    for (var arg: Type.getArgumentTypes(mc.methodnode.desc))
        insert(new InsnNode(arg.getSize() == 1 ? Opcodes.POP : Opcodes.POP2);
        
    remove(mc.ins);
}
\end{lstlisting}
\subsection{Code Analysis of Programs}
\label{sec:codeanalysis}
%
%
We consider the analysis of program codes along quality metrics on class files.
Software quality is a classic concern in software engineering.
Measuring software quality is instrumental in ensuring several properties such as low technical debt, upgradable software, and secure coding. 
In~\cite{HongleiMetrics,AllenMetrics}, white-box (i.e., based on source code) analysis metrics are defined to measure quality, understandability, and maintainability.
The higher level of abstraction and the updatability of the source code (access to the documentation, commentaries, fully structured\ldots) are incentives for defining code analysis techniques on source code.
As such, there is a lack of tools to compute quality metrics on the bytecode.

BISM permits access and compute many valuable properties that can be used to compute standard metrics relying on the CFG of methods, the number of variables and method calls, and the program instructions.
This makes such analysis possible on legacy software.

While BISM does not provide access to the source code nor to some classical metrics like Lines Of Code or NPATH complexity, it still provides essential static information.
Next, we show how to compute the following software quality metrics: Mc Cabe Cyclomatic complexity, ABC Metric, and the count of unused variables.
%
\paragraph*{Mc Cabe complexity.}
%
The Mc Cabe Cyclomatic complexity~\cite{mccabe} is defined as the maximum number of independent paths in a CFG.
For a CFG $G$, it is easily computable by: $V(G) = |\mathit{Edges}_G| - |\mathit{Nodes}_G| + 2$.
%
%
In Listing~\ref{mccabe}, the transformer uses the computed CFG to count the number of conditional edges inside it.
\begin{lstlisting}[style=UseCase, caption={Mc Cabe cyclomatic complexity.}, label=mccabe]
int edgeNumber;
public void onMethodEnter(...){
    edgeNumber = 0;
}
public void onBasicBlockExit(BasicBlock bb,...){
    switch (bb.blockType){
        case CONDJUMP:
        case SWITCH:
         edgeNumber++;
    }
}
public void onMethodExit(Method m,...) {
    int c = m.getNumberOfBasicBlocks() - edgeNumber + 2;
    Log("Cyclomatic complexity of "+m.name+" is : "+ c);
}
\end{lstlisting}
%
\paragraph*{ABC Complexity.}
%
To compute the ABC complexity, we only need to classify instructions and basic blocks.
Computing the ABC complexity~\cite{abc} relies on the capability to distinguish between branching, assignments, and conditional jumps.
%
The transformer does it in Listing~\ref{ABC} using \inlineJava{blockType} and \inlineJava{opcode} fields of static contexts.

\begin{lstlisting}[style=UseCase, caption={ABC complexity.}, label=ABC]
public void onMethodEnter(Method m,...) {
    A=B=C=0;
    C = m.methodNode.tryCatchBlocks.size(); //To count the try and catch in conditional
}

public void beforeInstruction(Instruction ins,...) {
    if (isAssignInstr(ins))
        A++;
    //Handle branches
    if (ins.opcode == Opcodes.GOTO || ins.opcode == Opcodes.NEW || ins.isBranchingInstruction())
        B++;
}

public void beforeMethodCall(MethodCall mc,...) {
    B++;
}

public void onBasicBlockExit(BasicBlock bb,...) {
    if (bb.blockType == BlockType.CONDJUMP)
        C++;
}

public void onMethodExit(Method m,...) {
    Log("ABC of "+m.name+" is "+ Math.sqrt(A*A+B*B+C*C));
}
\end{lstlisting}
%
\paragraph*{Unused variables.}
%
We consider that a variable is not used in a method if it is never loaded within the method. 
For this, the transformer in Listing~\ref{unusedVars} checks whether an instruction is a \emph{direct load} which takes as parameter a variable index and pushes its value onto the stack.
%
In such a case, the transformer retrieves the index of the variable and sets it as \emph{not unused} in the mapping implemented by boolean array \inlineJava{unusedVars}.
The check is run on all instructions, and the variables which have never been loaded on the stack are declared unused.

\begin{lstlisting}[style=UseCase, caption={Unused variables.}, label=unusedVars]
public void onMethodEnter(Method m,...) {
    unusedVars = new boolean[ m.methodNode.localVariables.size()];
    Arrays.fill(unusedVars, true);
}

public void beforeInstruction(Instruction ins,...) {
    if (ins.opcode >= Opcodes.LDC && ins.opcode <= Opcodes.SALOAD){
        //Loading a local variable, therefore variable is used
        if (ins.node instanceof VarInsnNode)
            unusedVars[((VarInsnNode) ins.node).var] = false;
        else if (ins.node instanceof IincInsnNode)
            unusedVars[((IincInsnNode) ins.node).var] = false;
    }
}
\end{lstlisting}
\subsection{Good Java Practices}
%
This section demonstrates how to use BISM to instrument the code for monitoring classical runtime verification properties. 
%
We do not discuss the monitor and assume that it is implemented in a separate library.
%
\subsubsection{HasNext Property}
\label{sec:has-next} 
%
We consider the \textbf{HasNext} property on iterators which specifies that the \inlineJava{hasNext()} method should be called and return true before calling the \inlineJava{next()} method on an iterator.
This property 
Listing~\ref{lst:hasNext} shows a BISM transformer for monitoring the property.
We use the method call joinpoints and filter for invocations of \inlineJava{hasNext()} and \inlineJava{next()} on iterator objects using the \inlineJava{mc} object which exposes static context from the captured method call.
We use the dynamic context object \inlineJava{dc} to retrieve the object receiving the method call, in this case, the \inlineJava{Iterator} instance.
The \inlineJava{getMethodReceiver} (explained in Section~\ref{sec:dynamiccontext}) retrieves the iterator instance by loading it from the stack into a local variable returning a reference to it in a 
\inlineJava{DynamicValue} object. 
Assuming that a monitor is implemented in a separate class with two static methods \inlineJava{hasNext()} and \inlineJava{next()}.

We invoke each method, respectively passing the iterator instance to the monitor using the BISM invocation helper method \inlineJava{StaticInvocation}.
\begin{lstlisting}[style=UseCase, 
 caption={HasNext instrumentation.},
 label=lst:hasNext,
 ]
public void afterMethodCall(MethodCall mc, MethodCallDynamicContext dc) {
    if (mc.methodName.contains("hasNext") && mc.methodOwner.contains("Iterator")){

        DynamicValue iterator = dc.getMethodReceiver(mc); //Instance of the iterator
        DynamicValue result = dc.getMethodResult(mc);

        StaticInvocation sti = new StaticInvocation("Monitor", "hasNext");
        sti.addParameter(iterator);
        sti.addParameter(result);
        invoke(sti);
    }
}
public void afterMethodCall(MethodCall mc, MethodCallDynamicContext dc) {
    if (mc.methodName.contains("next") && mc.methodOwner.contains("Iterator")){

        DynamicValue iterator = dc.getMethodReceiver(mc);
        StaticInvocation sti = new StaticInvocation("Monitor", "next");
        sti.addParameter(iterator);
        invoke(sti);
    }
}
\end{lstlisting}

\subsubsection{Safe Locking}
\label{sec:safe-lock} 

The \textbf{SafeUnlock} property specifies that the number of acquires and releases of a (reentrant) \inlineJava{Lock} class are matched within a given method call.
In Listing \ref{lst:safeLock}, the transformer captures lock and unlock operations in a method and extracts dynamic context such as the thread name, the lock object, the calling object instance.
A monitor is implemented in a separate class with the two static methods \inlineJava{lockOperation()} and \inlineJava{unLockOperation()}.
We invoke each method, respectively passing the extracted values to the monitor.

\begin{lstlisting}[style=UseCase, 
 caption={SafeLock instrumentation.},
 label=lst:safeLock,
 ]
@Override
public void beforeMethodCall(MethodCall mc, MethodCallDynamicContext dc) {
 
    if (mc.methodName.equals("lock") && mc.methodOwner.contains("Lock")) {

        DynamicValue threadName = dc.getThreadName(mc);
        DynamicValue lockObject = dc.getMethodReceiver(mc); //Instance of the Lock
        DynamicValue _this = dc.getThis(mc);
        String currentMethod = mc.ins.methodName;

        StaticInvocation sti = new StaticInvocation("Monitor", "lockOperation");
        sti.addParameter(threadName);
        sti.addParameter(lockObject);
        sti.addParameter(_this);
        sti.addParameter(currentMethod);

        invoke(sti);

    }
}
@Override
public void afterMethodCall(MethodCall mc, MethodCallDynamicContext dc) {

    if (mc.methodName.equals("unlock") && mc.methodOwner.contains("Lock")) {

        DynamicValue threadName = dc.getThreadName(mc);
        DynamicValue lockObject = dc.getMethodReceiver(mc); //Instance of the Lock
        DynamicValue _this = dc.getThis(mc);
        String currentMethod = mc.ins.methodName;

        StaticInvocation sti = new StaticInvocation("Monitor", "unLockOperation");
        sti.addParameter(threadName);
        sti.addParameter(lockObject);
        sti.addParameter(_this);
        sti.addParameter(currentMethod);
        invoke(sti);
    }
}
\end{lstlisting}
\subsection{Dynamic Profiling}
\label{sec:profiling} 
 %
 %
We demonstrate how to implement dynamic profiling with BISM.
We collect dynamic context from a running program, including the number of method invocations, runtime types of method arguments (\secref{sec:callGraph}), number of allocated objects (\secref{sec:objectAllocation}), and return types (\secref{sec:returnTypes}).
We do not focus on implementing the profiler tool but only on how to extract context using BISM.
%

%
\subsubsection{Call Graph}
\label{sec:callGraph}
%
We consider the dynamic call graph of a program which represents the calling relationship between methods in program execution.
For each method call in an execution, we are interested in extracting runtime information from the calling and called methods.
Listing \ref{lst:callGraph} shows the code of a transformer that instruments to extract the \textit{caller} and \emph{callee} classes and method names along with their runtime arguments, at each method call.
%
The arguments of the caller and callee are extracted using the dynamic context method \inlineJava{dc.getMethodArgs()}.
We instrument two synthetic local arrays in the base program to store the extracted values locally in the method.
For the caller, the arguments are retrieved once at method enter to avoid repeating the argument extraction for each invocation by the caller.
At \inlineJava{onMethodEnter}, the \inlineJava{dc.getMethodArgs()} will retrieve the needed values from the local variables of the method.
As for the callee, the \inlineJava{dc.getMethodArgs()} will retrieve the arguments directly from the stack.
Then, before each method call, an invocation to the profiler method \inlineJava{callGraph} is instrumented, passing the static and dynamic information.
\begin{lstlisting}[style=UseCase,
 caption={Profiling the call graph.},
 label=lst:callGraph,
 ]
LocalArray callerArgs;
LocalArray calleeArgs;

public void onMethodEnter(Method m, MethodDynamicContext dc){
    //Initialize the local arrays
    callerArgs = dc.createLocalArray(m);
    calleeArgs = dc.createLocalArray(m);

    int args = m.getNumberOfArguments();
    DynamicValue dv;
    for (int i = 1; i < args + 1; i++) {
        dv = dc.getMethodArgs(m, i);
        dc.addToLocalArray(callerArgs, dv);
    }
}

public void beforeMethodCall(MethodCall mc, MethodCallDynamicContext dc){
    dc.clearLocalArray(mc, calleeArgs);

    int args = mc.getNumberOfArgs();
    DynamicValue dv;
    for (int i = 1; i < args + 1; i++) {
        dv = dc.getMethodArgs(mc, i);
        dc.addToLocalArray(calleeArgs, dv);
    }

    //Invoke profiler 
    StaticInvocation sti = new StaticInvocation("Profiler", "callGraph");
    sti.addParameter(dc.getThreadName(mc));
    sti.addParameter(mc.method.fullName);
    sti.addParameter(callerArgs);
    sti.addParameter(mc.fullName);
    sti.addParameter(calleeArgs);
    invoke(sti);
}
\end{lstlisting}
%
\subsubsection{Object Allocation}
\label{sec:objectAllocation} 
%
Object allocation is an important metric in dynamic profiling that allows the user to know the number of created objects in the program and estimate the used memory.
Listing \ref{lst:objectAllocation} shows a transformer that instruments to capture allocated objects and arrays in a program.
We use the \inlineJava{beforeInstruction} joinpoint and filter for all \inlineJava{NEW} opcodes.
To extract the type of the created object, we use the access granted by BISM to the ASM instruction node object and get more details from the bytecode instruction.
The extracted static information is then passed to the profiler by invoking its appropriate method.
\begin{lstlisting}[style=UseCase,
 caption={Profiling object allocation.},
 label=lst:objectAllocation
 ]
@Override
public void beforeInstruction(Instruction ins,...) {
    //Object creation opcodes
    if (ins.opcode == Opcodes.NEW
    || ins.opcode == Opcodes.NEWARRAY
    || ins.opcode == Opcodes.ANEWARRAY
    || ins.opcode == Opcodes.MULTIANEWARRAY) {

        TypeInsnNode instruction = (TypeInsnNode) ins.node;
        //Invoke profiler
        StaticInvocation sti = new StaticInvocation("Profiler", "allocation");
        sti.addParameter(ins.method.fullName);
        sti.addParameter(ins.opcode);
        sti.addParameter(instruction.desc);
        invoke(sti);
    }
}
\end{lstlisting}
%
\subsubsection{Return Types}
\label{sec:returnTypes} 
%
Listing \ref{lst:returnTypes} shows how instrument to extract return types from methods.
We use the \inlineJava{afterMethodCall} joinpoint and filter using the static context provided \inlineJava{mc.returns} which returns a boolean flag indicating if the method has a return type in its signature.
Then, we extract the return result into the dynamic value object \inlineJava{dv}.
After that, an invocation to the profiler is instrumented, which passes the needed information.
We choose to box the return value for a more generic implementation.

\begin{lstlisting}[style=UseCase,
 caption={Profiling return types.},
 label=lst:returnTypes,
 ]
 @Override
public void afterMethodCall(MethodCall mc, MethodCallDynamicContext dc){
    //If a method returns 
    if (mc.returns) {
        //Get the result
        DynamicValue dv = dc.getMethodResult(mc); 

        //Invoke profiler
        StaticInvocation sti = new StaticInvocation("Profiler", "returnTypes");
        sti.addParameter(caller);
        sti.addParameter(mc.fullName);
        sti.addBoxedParameter(dv);
        invoke(sti);
    }
}
\end{lstlisting}
\subsection{Logging}
\label{sec:logging} 
%
%
Logging is a classic example of a cross-cutting concern that is better implemented following the aspect-oriented paradigm.
Indeed, by using Java annotations, one can mark methods that require logging and avoid polluting the source code with multiple logging instructions.
This way, one can instrument the program and insert the logging instructions only on annotated methods.

Listing \ref{lst:logging} shows a transformer that instruments the program to log the execution of selected methods in a program on method entries and exits.
One way to mark methods that need to be logged in an application by a developer is by creating a custom annotation and annotating the needed methods.
The transformer looks for a hypothetical \inlineJava{@Log} annotation inserted at methods in the base program.
The annotation indicates that the method needs to be logged.
BISM provides access to annotations on methods through the static context.
The \inlineJava{isAnnotated(Log)} returns a boolean flag that indicates if the method is annotated with \inlineJava{@Log}.
Then, a simple log message is printed on the console.
This example can be extended to extract and log the arguments passed to the method (see \secref{sec:callGraph}).

\begin{lstlisting}[style=UseCase, 
 caption={Logging.},
 label=lst:logging,
 ]
@Override
public void onMethodEnter(Method m, MethodDynamicContext dc){
    if (m.isAnnotated("Log")) //Checks annotation on method
        println("Entering method: " + m.name);
}

@Override
public void onMethodExit(Method m, MethodDynamicContext dc){
    if (m.isAnnotated("Log"))
        println("Exiting method: " + m.name);
}

\end{lstlisting}

%
\section{BISM Implementation}
\label{sec:framework}

\begin{figure*}
\centering
 \includegraphics[width=0.9\textwidth]{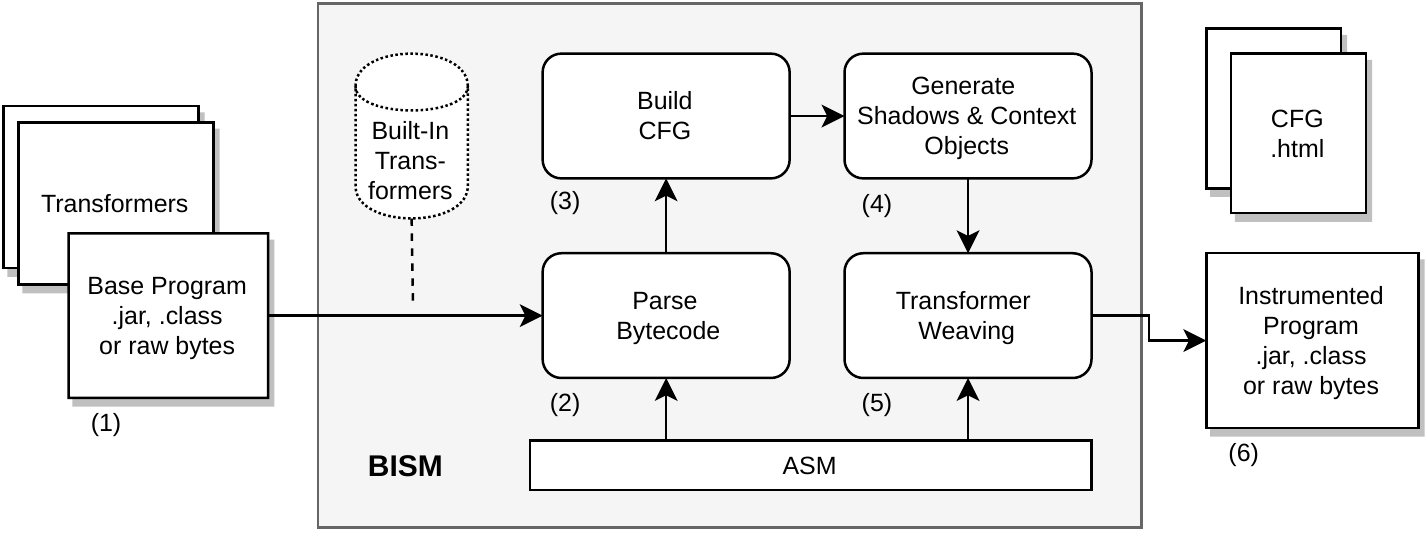} 
 \caption{Instrumentation process in BISM.} 
 \label{fig:framework-overview}
\end{figure*}

In this section, we provide some details about BISM implementation. 
BISM~\cite{bism} is implemented in Java using about 7,000 LOC and 55 classes distributed in separate modules. 
It uses ASM for bytecode parsing, analysis, and weaving. 
BISM can run in two modes: a build-time mode where BISM runs as a standalone application to statically instrument a program, and a load-time mode where BISM is attached to a program as a Java agent. 
Fig.~\ref{fig:framework-overview} shows BISM internal workflow.\\[1ex]
%
%
%

\textbf{(1) User Input.}
In build-time mode, BISM takes a base program bytecode (\emph{.class} or \emph{.jar}) to be instrumented and a list of transformers that specifies the instrumentation logic. 
In load-time mode, BISM only takes the transformers and instruments all classes being loaded by the JVM. 
BISM provides several built-in transformers that can be directly used. 
Moreover, users can specify various runtime arguments to BISM or even the transformers, from the console or through a configuration file.\\[1ex]
\textbf{(2) Parse Bytecode.}
For each class in the base program, BISM uses ASM to parse the bytecode and generate a tree object containing all the class details, such as fields, methods, and instructions. 
The following three steps will be performed on each class for every transformer specified in a run.\\[1ex]
%
\textbf{(3) Build CFG.}
BISM constructs the CFGs for all methods in the target class. 
If the transformer utilizes control-flow joinpoints (\ttt{onTrueBranch} and \ttt{onFalseBranch}), BISM eliminates all \emph{critical edges} from the CFGs to avoid instrumentation errors. 
This is done by inserting empty basic blocks in the middle of critical edges, which is only applied if used while keeping copies of the original CFGs.

Also, if the transformer uses joinpoint \ttt{onMethodExit}, all the exit blocks (which terminate with a return opcode) are merged into a single to avoid duplication and errors.
This is done by adding a new block that contains a return of a suitable type; then, all other returns are replaced by unconditional jumps to the added one.
Moreover, if the users opted for the \emph{visualizer}, the CFGs are printed into HTML files on the disk. \\[1ex]
\textbf{(4) Generate Shadows and Context Objects.}
%
BISM iterates over the target class to identify all shadows utilizing the created CFGs.
The relevant static and dynamic context objects are created and initialized using the static information available and BISM analysis at each shadow. \\[1ex] 
%
\textbf{(5) Transformer Weaving.}
The transformer is notified of each shadow and passed the static and dynamic objects.
The weaving loop is illustrated in Figure~\ref{fig:instr-loop}.
BISM evaluates the transformations applied by a transformer using the advice methods.
After that, it accordingly weaves the necessary bytecode instructions into the target class. \\[1ex]
%
\textbf{(6) Output.}
The instrumented bytecode is then output back as a \emph{.class} file in build-time mode or passed as raw bytes to the JVM in load-time mode. 
In case of instrumentation errors, e.g., due to adding manual ASM instructions, BISM emits a weaving error. 
If the \emph{visualizer} is enabled, the instrumented CFGs are also printed into HTML files on the disk. 
%
%
\section{Performance Evaluation}
\label{sec:casestudies}
We report on our performance evaluation of BISM.
\paragraph{Experiments and used programs.}
We compare BISM with DiSL and AspectJ, which are the most popular tools for the monitoring and runtime verification of Java programs.
For this, we use three complementary experiments\footnote{We use the latest versions of DiSL from \url{https://gitlab.ow2.org/disl/disl} and AspectJ Weaver 1.9.4.}. Table~\ref{tab:comp} illustrates how the three experiments are complementary to each other. 
\begin{itemize}
\item 
The first experiment concerns the implementation of the Advanced Encryption Standard (AES).
This experiment shows how BISM can perform inline instrumentation by inserting new bytecode instructions inside the target program to detect test inversion attacks on the application.
\item 
The second experiment concerns a financial transaction system.
This experiment shows how BISM can be used to instrument the system to monitor user-provided properties.
The financial transaction system is a relatively small application with a low event rate.
\item 
The third experiment concerns the DaCapo benchmark~\cite{DaCapo06}.
This experiment shows how BISM can be used to instrument the benchmark and monitor for the good usage of data structures (with classical properties \textbf{HasNext}, \textbf{UnSafeIterator}, and \textbf{SafeSyncMap}).
DaCapo is a large benchmark classically used when evaluating runtime verification tools as it produces events at a high rate.
\end{itemize}
For the first experiment, the instrumentation is performed at the level of the control-flow graph.
For the two other experiments, the instrumentation is performed at the level of method calls to emit events.
Note, AspectJ is not capable of instrumenting for inline monitoring of control-flow events, so we do not include it in the first experiment (with AES).

We run our experiments in both BISM instrumentation modes, namely load-time and build-time.
Running an experiment in load-time mode serves to compare the performance when the tools act as an interface between the base program and the virtual machine.
Running an experiment in build-time mode serves to compare the performance of the generated instrumented bytecode.

We note that DiSL wraps its instrumentation code with exception handlers.
Exception handlers are not necessary for our experiments and have a performance impact.
To guarantee fairness, we switched off exception handlers in DiSL.
%
%
\paragraph{Evaluation metrics.}
We consider three performance metrics: runtime, used memory, and bytecode-size.
We are interested in evaluating the instrumentation overhead, that is, the performance degradation caused by instrumentation.
For each metric, we use the base program as a baseline.
For runtime, we measure the execution time of the instrumented program.
For used memory, we measure the used heap and non-heap memory after a forced garbage collection. 
In load-time mode, we do not measure the used memory in the case of DiSL because DiSL performs instrumentation on a separate JVM process. 
\paragraph{Evaluation environment.}
To run the experiments, we use Java JDK 8u251 with \SI{2}{\giga\byte} maximum heap size on an Intel Core i9-9980HK (2.4 GHz. \SI{8}{\giga\byte} RAM) running Ubuntu 20.04 LTS 64-bit.
We consider 100 runs and then calculate the mean and the standard deviation. 

In what follows, we illustrate how we carried out our experiments and the obtained results\footnote{
More details about the experiments and the material needed to reproduce them can be found at \url{https://gitlab.inria.fr/bism/bism-experiments}. 
}. 

\bgroup
\begin{table*}[htbp]
 \centering
 \caption{A comparison between the experiments. LT is for load-time mode, and BT is for build-time mode. A checkmark (\dgreen\cmark) indicates that the experiment involves the metric or the feature, whereas a cross mark (\red\xmark) indicates that the experiment does not involve the metric or the feature. Term NA abbreviates Not Applicable, and (-DiSL) indicates that the DiSL tool has been excluded.}
 \renewcommand{\arraystretch}{1.2}
 \setlength{\tabcolsep}{3pt}
 \begin{tabular}{ccccccccc}
 \toprule
&&\multicolumn{3}{c}{\bf \hspace{-15pt}Performance Metrics} & \bf Instrumentation & \bf Bytecode & \multicolumn{2}{c}{\bf Comparison with} \\
\multicolumn{2}{c}{~} 
& \textbf{Runtime} & \textbf{Used Memory} & \textbf{Bytecode Size} & \textbf{Level} & \textbf{Insertion} & \textbf{AspectJ} & \textbf{DiSL} \\
 \midrule
 \multirow{2}{*}{AES} 
 & LT & \both{\dgreen\cmark} & \dgreen\cmark ~(-DiSL) & NA & \both{Low (CFG-Level)} & \both{\dgreen\cmark} & \both{NA} & \both{\dgreen\cmark} \\
 & BT & & \dgreen\cmark & \dgreen\cmark & & & & \\
 \midrule
 \multirow{2}{*}{Transactions} 
 & LT & \both{\dgreen\cmark} & \dgreen\cmark ~(-DiSL) & NA & \both{High (Method Calls)} & \both{\red\xmark} & \both{\dgreen\cmark} & \both{\dgreen\cmark} \\
 & BT & & \dgreen\cmark & \dgreen\cmark & & & & \\
 \midrule
 \multirow{2}{*}{DaCapo} 
 & LT & \both{\dgreen\cmark} & \dgreen\cmark ~(-DiSL) & NA & \both{High (Method Calls)} & \both{\red\xmark} & \both{\dgreen\cmark} & \both{\dgreen\cmark} \\
 & BT & & \dgreen\cmark & \dgreen\cmark & & & & \\
 \bottomrule
 \end{tabular}
 \label{tab:comp}
 \end{table*}
\egroup

\subsection{Advanced Encryption Standard (AES)}
\label{sec:aes}
\paragraph{Experimental setup.}
We compare BISM with DiSL in a scenario using inline monitors.
We instrument an external AES implementation to detect test inversions in the control flow of the program execution. 
The instrumentation deploys inline monitors that duplicate all conditional jumps in their successor blocks to report test inversions.
We implement the instrumentation as follows. 

In BISM, we use built-in features to duplicate all conditional jumps utilizing the ability to insert raw bytecode instructions.
In particular, we use the instrumentation locator \inlineJava{beforeInstruction} to capture conditional jumps.
To extract the opcode for each conditional jump, we use the static context object \inlineJava{Instruction}, and to duplicate the operand values on the stack,
 we use the advice method \inlineJava{insert}\footnote{Extracting stack values can be also alternatively achieved using dynamic context method \inlineJava{getStackValue} and adding new local variables.}. 
%
%
We then use the control-flow instrumentation locators\footnote{ \inlineJava{OnTrueBranchEnter,onFalseBranchEnter}.} to capture the successor blocks executing after every conditional jump.
Finally, at the beginning of these blocks, we utilize \inlineJava{insert} to duplicate the conditional jump instruction. 

In DiSL, we implement a custom \inlineJava{StaticContext} object to retrieve information from conditional jump instructions, such as the indices of jump targets and instruction opcodes. 
Note, we use multiple \inlineJava{BytecodeMarker} snippets to capture all conditional jumps.
To retrieve stack values, we use the dynamic context object.
We then store the extracted information in synthetic local variables, and we add a flag to specify that a jump has occurred. 
Finally, on successor blocks, we map opcodes to Java syntax to re-evaluate conditional jumps using switch statements.
%

\paragraph{Load-time evaluation.}
%

\begin{figure}[htbp]
 \centering
 \includegraphics[width=0.5\textwidth]{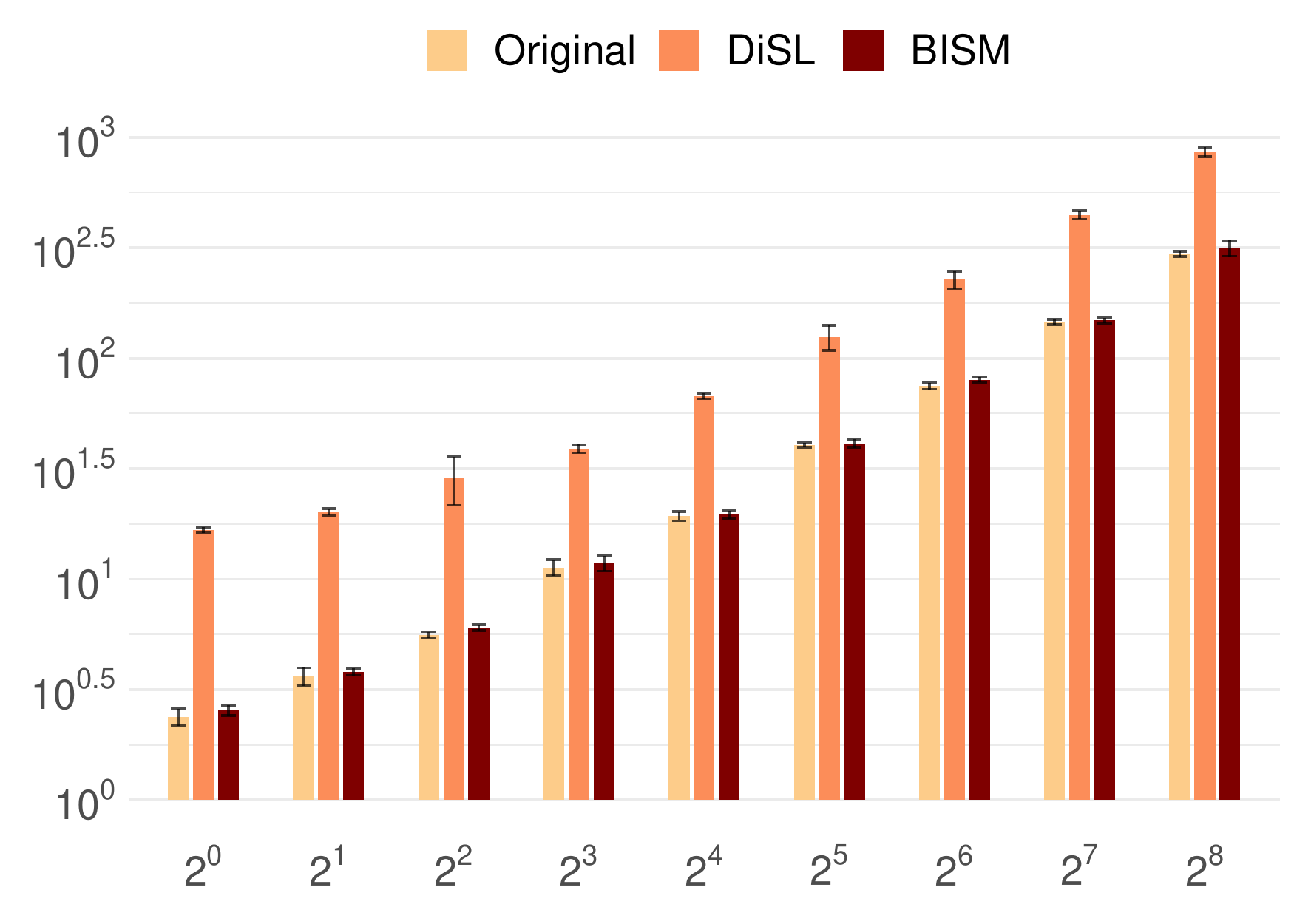} 
 \caption{AES load-time instrumentation runtime (ms).}
 \label{fig:aesloadtime}
\end{figure}

We consider different sizes of the plain text to be encrypted by AES. 
Figure~\ref{fig:aesloadtime} reports runtime with respect to plain-text size, in load-time mode.
BISM shows better performance over DiSL for all plain-text sizes. 
We do not measure the used memory because DiSL performs instrumentation on a separate JVM process which imposes a huge memory overhead.
Also, AspectJ is excluded from this experiment as it cannot capture control-flow events.

\paragraph{Build-time evaluation.}
\begin{figure*}[t]
\centering
 \begin{subfigure}[b]{0.495\textwidth}
 \includegraphics[width=\textwidth]{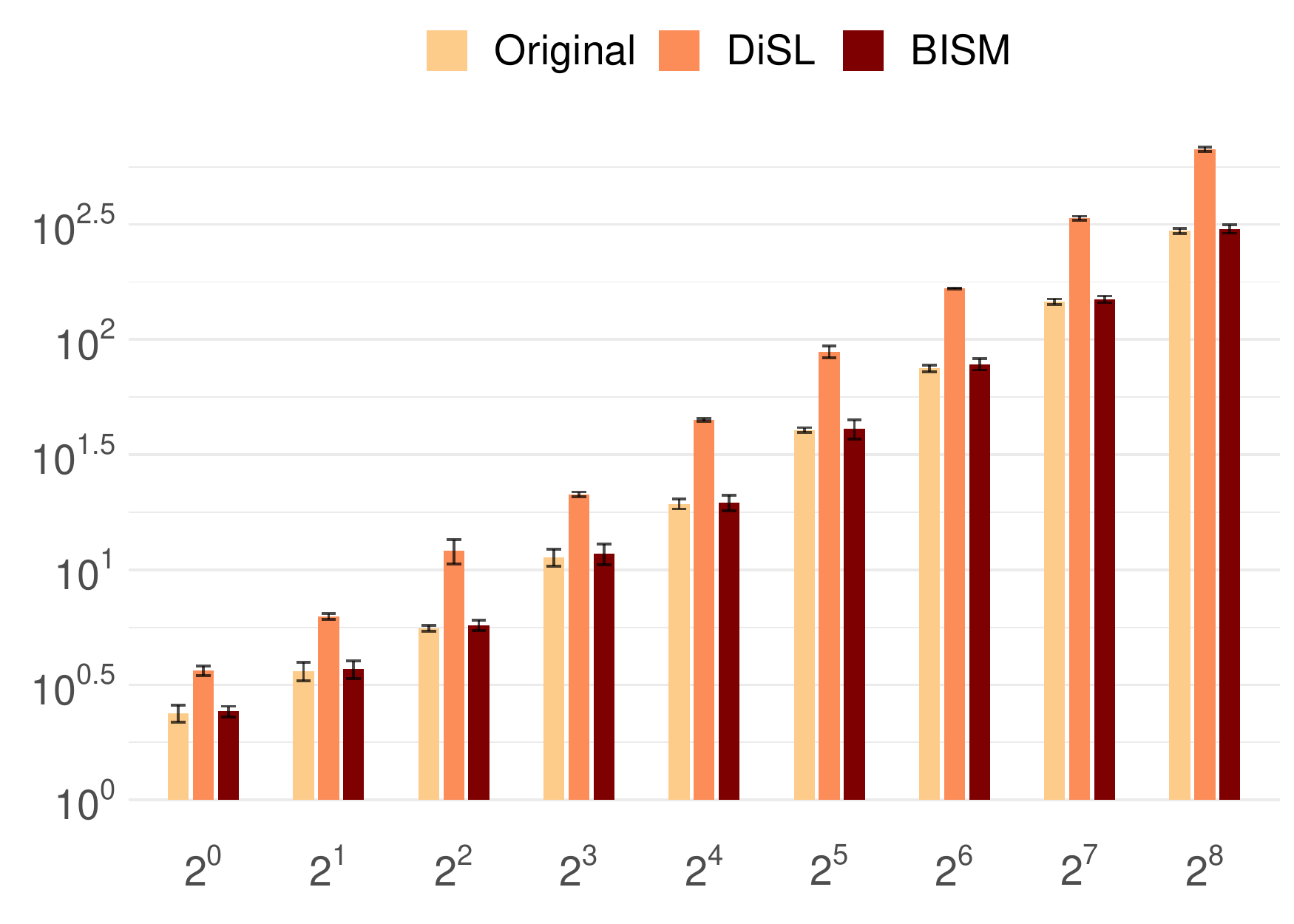}
 \caption{Runtime (ms).}
 \end{subfigure}
 \hfill
 \begin{subfigure}[b]{0.495\textwidth}
 \includegraphics[width=\textwidth]{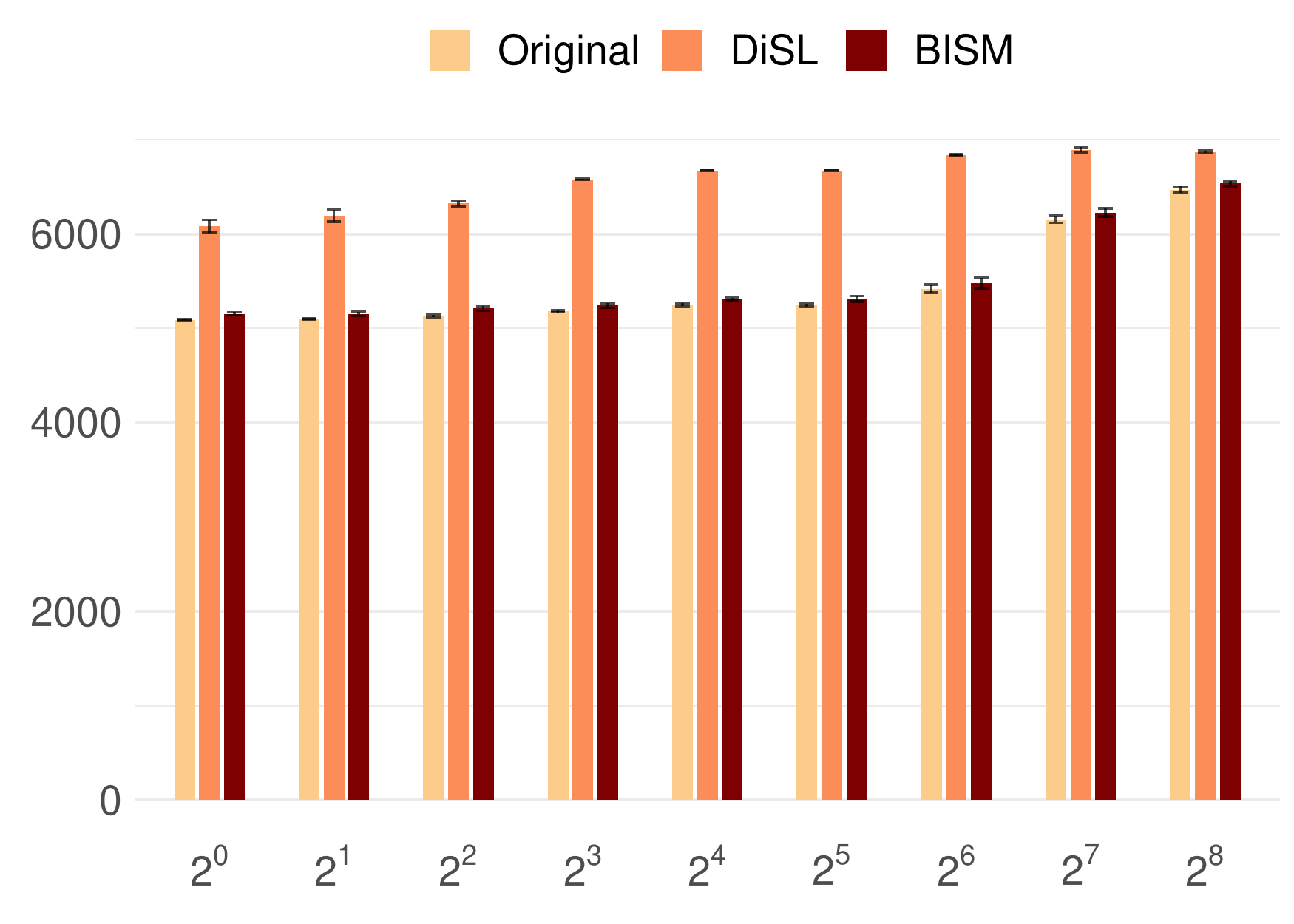}
 \caption{Used memory (\si{\kilo\byte}).}
 \end{subfigure}
 \caption{AES build-time instrumentation.}
 \label{fig:aesbuildtime}
\end{figure*}

We replace the original classes of AES with statically instrumented classes from each tool.
Figure~\ref{fig:aesbuildtime} reports the runtime and used memory for plain-text size in build-time mode.
%
%
BISM shows less overhead than DiSL in both runtime and used memory for all plain-text sizes. 
Moreover, BISM incurs a relatively small overhead for all plain-text sizes. 
%
%
Table~\ref{tab:aes} reports the number of generated events (corresponding to conditional jumps) after running the code (in millions).
\bgroup
\setlength{\tabcolsep}{3pt}
\begin{table}[]
\caption{Number of emitted events in AES experiment.}
\label{tab:aes}
 \centering
 \begin{tabular}{cccccccccc} 
 %
 \toprule
 \textbf{Plain-text size (\si{\kilo\byte})} & $2^0$ & $2^1$ & $2^2$ & $2^3$ & $2^4$ & $2^5$ & $2^6$ & $2^7$ & $2^8$\\
 \midrule
 \textbf{Events (\si{\mega})} & 0.9 & 1.8 & 3.6 & 7.3 & 14.9 & 29.5 & 58.5 & 117 & 233\\ 
 \bottomrule
 \end{tabular}
\end{table}
\egroup
The bytecode size of the original AES class is \SI{9}{\kilo\byte}. 
After instrumentation, the bytecode size is \SI{10}{\kilo\byte} (+11.11\%) for BISM, and \SI{128}{\kilo\byte} (+1,322\%) for DiSL. 
So, BISM incurs less bytecode-size overhead than DiSL. 
The significant overhead in DiSL is due to the inability to inline the monitor in bytecode and having to instrument it in Java.
We note that it is not straightforward in DiSL to extract control-flow information in Markers, whereas BISM provides this out-of-the-box.
\subsection{Financial Transaction System}
\label{sec:transactions}

\paragraph{Experimental setup.}
We compare BISM with DiSL and AspectJ in a runtime verification scenario to monitor some properties of a financial transaction system. 
We use the implementation from CRV-14~\cite{BartocciFBCDHJK19} to monitor the following properties: 
\begin{itemize}
 \item
 Property P1: only users based in certain countries can be Silver or Gold users. 
 \item
 Property P2: the transaction system must be initialized before any user logs in.
 \item
 Property P3: no account may end up with a negative balance after being accessed.
 \item
 Property P4: an account approved by the administrator may not have the same account number as any other already existing account in the system.
 \item
 Property P5: once a user is disabled by the administrator, he or she may not withdraw from an account until being activated again by the administrator.
 \item
 Property P6: once greylisted, a user must perform at least three deposits from external before being whitelisted.
 \item
 Property P7: no user may request more than 10 new accounts in a single session.
 \item
 Property P8: the administrator must reconcile accounts every 1000 external transfers or an aggregate total of one million dollars in external transfers.
 \item
 Property P9: a user may not have more than three active sessions at once.
 \item
 Property P10: transfers may only be made during an active session (i.e., between a login and logout). 
\end{itemize} 
For each property using a set of events, we instrument the financial transaction system to generate those events.
Such events mainly correspond on the system to method call with parameters or class field updates.
For example, monitoring Property P6 requires the following events: \ttt{greylistUser(id)}, \ttt{depositFromExternal(id)} and \ttt{whitelistUser(id)}, where \ttt{id} is a unique user identifier. 
We implement a set of related scenarios and an external monitor library with stub methods that only count the number of received events.
We implement instrumentation as follows: 
\begin{itemize}
 \item 
 In BISM, we use the static context provided at method-call instrumentation \locators{}\footnote{\inlineJava{beforeMethodCall}, \inlineJava{afterMethodCall}.} to filter methods by their names and owners. 
 To access the method calls' receivers and results, we utilize methods \inlineJava{getMethodArgs} and \inlineJava{getMethodResult} available in dynamic contexts. 
 We then use argument processors and dynamic context objects to access dynamic values and pass them to the monitor.
 The extracted values are then passed to the monitor by invoking its appropriate method.
 \item
 In DiSL, we implement custom Markers to capture the needed method calls and use argument processors and dynamic context objects to access dynamic values. 
 We note that it required to create a custom marker for each method call, which resulted in implementing 28 different marker classes.
 \item
 In AspectJ, we use the call pointcut, type pattern matching, and joinpoint static information to capture method calls and write custom advices that invoke the monitor.
\end{itemize}
%

\paragraph{Load-time evaluation.}
\begin{figure*}[t]
\centering
 \begin{subfigure}[b]{0.495\textwidth}
 \includegraphics[width=\textwidth]{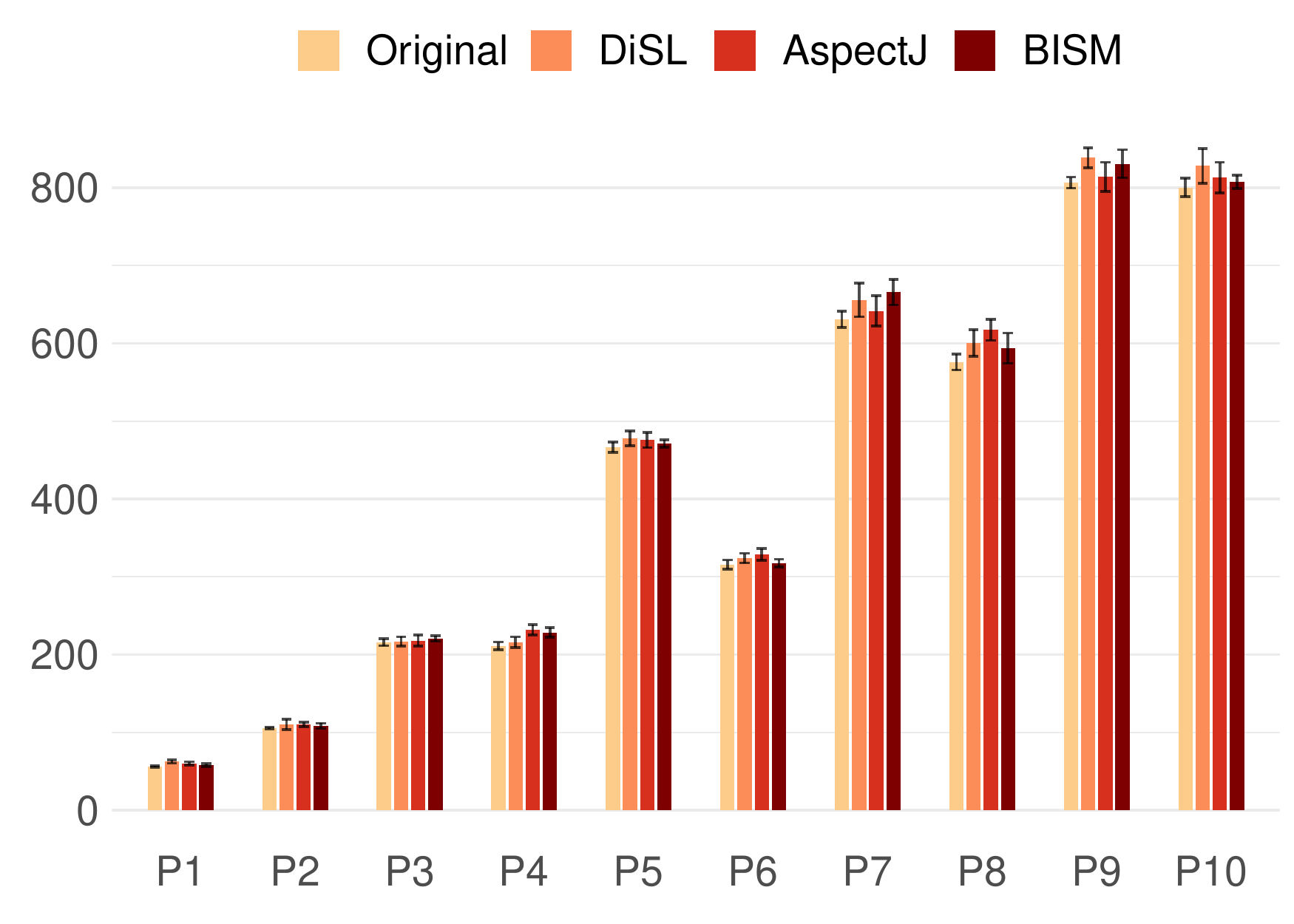}
 \caption{Runtime (ms).}
 \end{subfigure}
 \hfill
 \begin{subfigure}[b]{0.495\textwidth}
 \includegraphics[width=\textwidth]{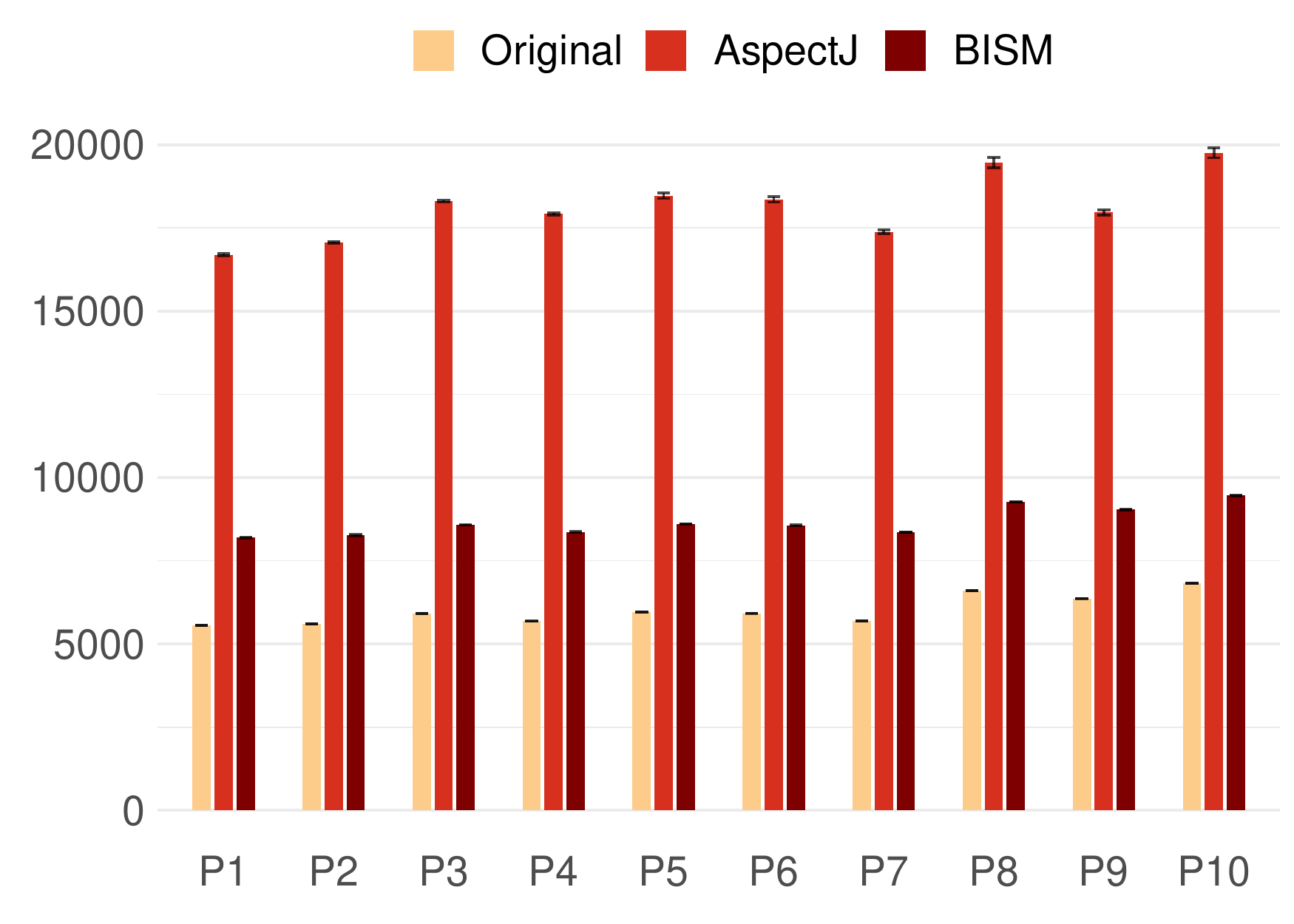}
 \caption{Used memory (\si{\kilo\byte}).}
 \end{subfigure}
 \caption{Financial transaction system load-time instrumentation.}
 \label{fig:transactionsloadtime}
\end{figure*}

Figure~\ref {fig:transactionsloadtime} reports the runtime and used memory for the considered properties in load-time mode (excluding DiSL in the case of used memory). 
%
%
%
BISM shows better performance over DiSL and AspectJ for properties P2, P5, P6, P8, and P10, for five properties out of ten.
Whereas DiSL shows the best performance for P3 and P4, and AspectJ shows the best performance for properties P1, P7, and P9.
The similar results of the tools is due to the fact that each property augments the base program with a small number of advices at limited locations, ranging between two and five advices per property.
Hence, the results in load-time mode reflect the execution time of the woven advice more than the instrumentation overhead.
Concerning used memory, BISM incurs much lower overhead than AspectJ for all properties. 
%

\paragraph{Build-time evaluation.}
\begin{figure*}[t]
\centering
 \begin{subfigure}[b]{0.495\textwidth}
 \includegraphics[width=\textwidth]{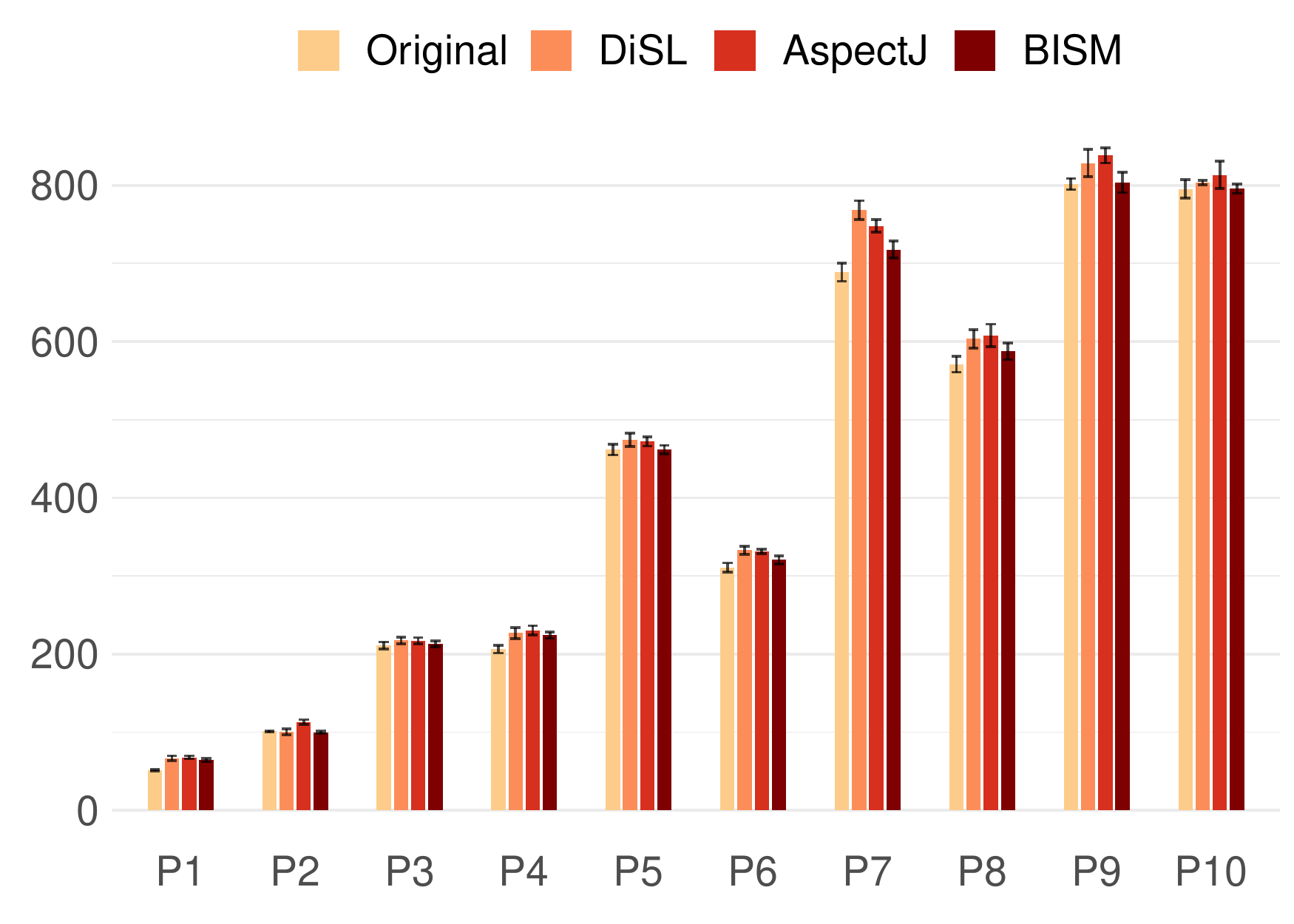}
 \caption{Runtime (ms).}
 \end{subfigure}
 \hfill
 \begin{subfigure}[b]{0.495\textwidth}
 \includegraphics[width=\textwidth]{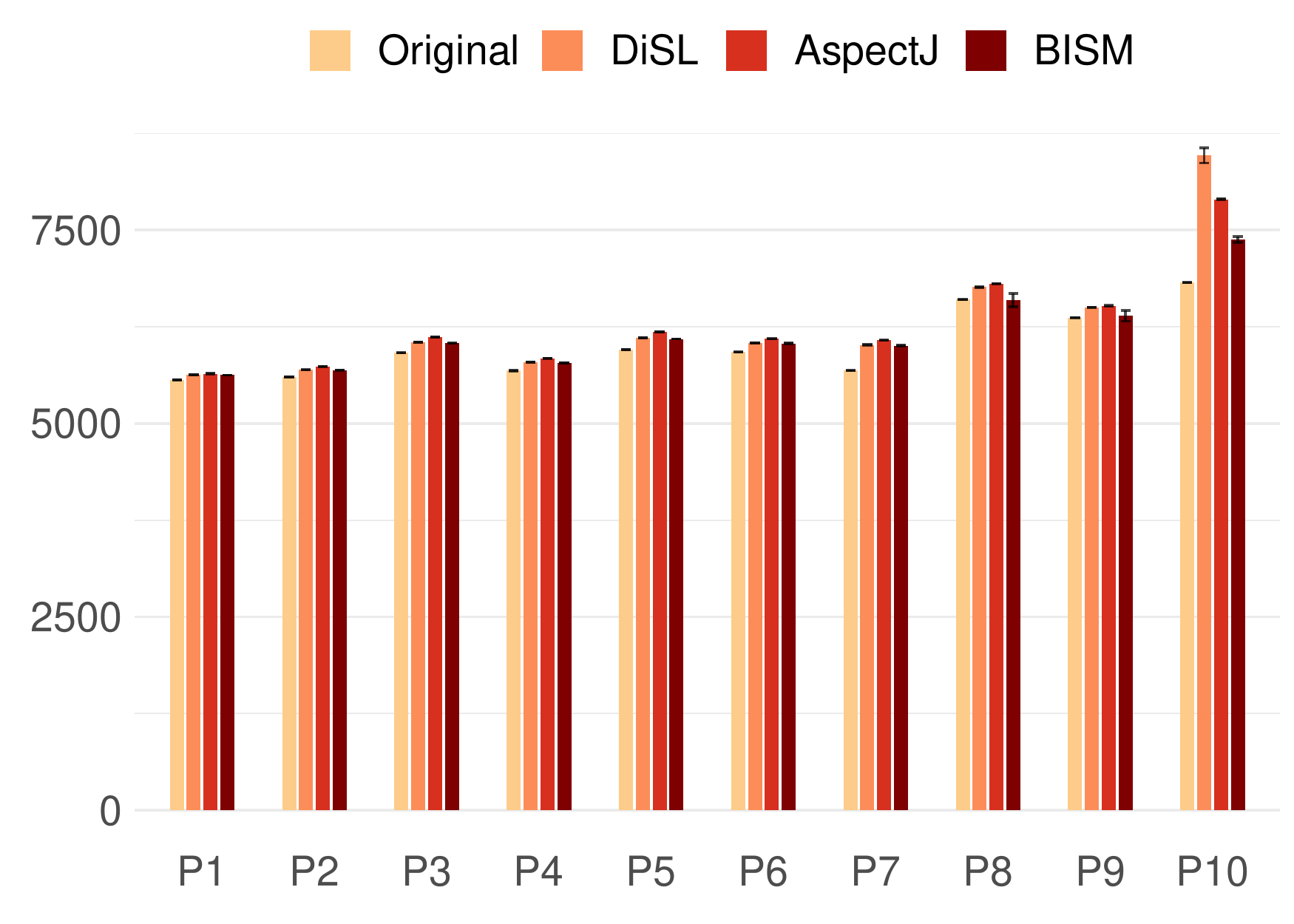}
 \caption{Used memory (\si{\kilo\byte}).}
 \end{subfigure}
 \caption{Financial transaction system build-time instrumentation.}
 \label{fig:transactionsbuildtime}
\end{figure*}

We replace the original classes of the scenarios with statically instrumented classes from each tool.
Figure~\ref {fig:transactionsbuildtime} reports the runtime and used memory for the considered properties in build-time mode.
BISM shows less runtime and used-memory overheads than both DiSL and AspectJ for all properties.
Table~\ref{tab:transactions} reports the number of generated events after running the code (in thousands).
The bytecode size of the classes of the overall original scenarios is \SI{44}{\kilo\byte}. 
After instrumentation, the bytecode size  is \SI{56}{\kilo\byte} (+27.27\%) for BISM, \SI{84}{\kilo\byte} (+90.9\%) for DiSL, and \SI{116}{\kilo\byte} (+163.63\%) for AspectJ.  
Hence, BISM incurs less bytecode-size overhead than both DiSL and AspectJ. 

\bgroup
\setlength{\tabcolsep}{4pt}
\begin{table}[]
\caption{Number of events generated by the financial transaction system for each monitored property.}
\label{tab:transactions}
 \centering
 \begin{tabular}{ccccccccccc} 
 \toprule 
 \textbf{Property} & P1 & P2 & P3 & P4 & P5 & P6 & P7 & P8 & P9 & P10 \\ 
 \midrule
 \textbf{Events} (k) & 10 & 10 & 11 & 33 & 71 & 69 & 125 & 104 & 192 & 154 \\
 \bottomrule
 \end{tabular}
\end{table}
\egroup

%
%
\subsection{DaCapo Benchmarks}
\label{sec:dacapo}
%
\paragraph{Experimental setup.}
We compare BISM with DiSL and AspectJ in a general runtime verification scenario.
We instrument the benchmarks in the DaCapo suite~\cite{DaCapo06} (dacapo-9.12-bach), to monitor for the classical \textbf{HasNext}, \textbf{UnSafeIterator}, and \textbf{SafeSyncMap} properties\footnote{The \textbf{HasNext} property specifies that a program should always call method \inlineJava{hasNext()} before calling method \inlineJava{next()} on an iterator.
 The \textbf{UnSafeIterator} property specifies that a collection should not be updated when an iterator associated with it is being used.
 The \textbf{SafeSyncMap} property specifies that a map should not be updated when an iterator associated with it is being used.}.
We only target the packages specific to each benchmark and do not limit our scope to \inlineJava{java.util} types; instead, we match freely by type and method name.
We implement an external monitor library with stub methods that only count the number of received events.
We implement the instrumentation similarly to the second experiment:
\begin{itemize}
 \item
In BISM, we use the static context provided at method-call instrumentation \locators{} to filter methods. 
%
%
\item
In DiSL, we implement custom Markers to capture the needed method calls. 
\item
In AspectJ, we use the call pointcut, type pattern matching, and joinpoint static information to capture method calls. 
\end{itemize}
%
\begin{figure}[htbp]
    \begin{subfigure}[b]{0.5\textwidth}
     \includegraphics[width=\textwidth]{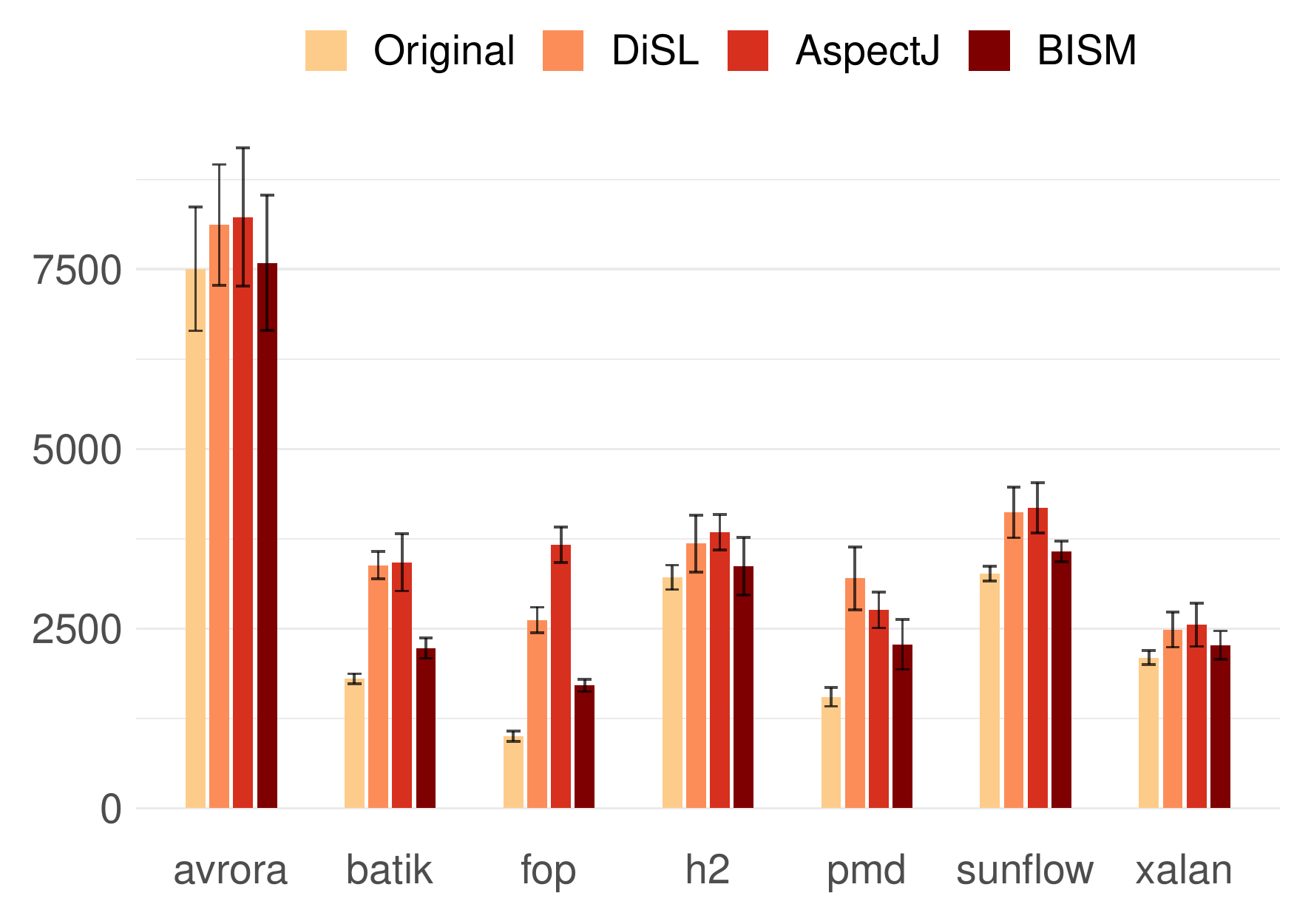} 
    \caption{Runtime (ms).}
    \end{subfigure}
    \hfill
    \begin{subfigure}[b]{0.5\textwidth}
    \includegraphics[width=\textwidth]{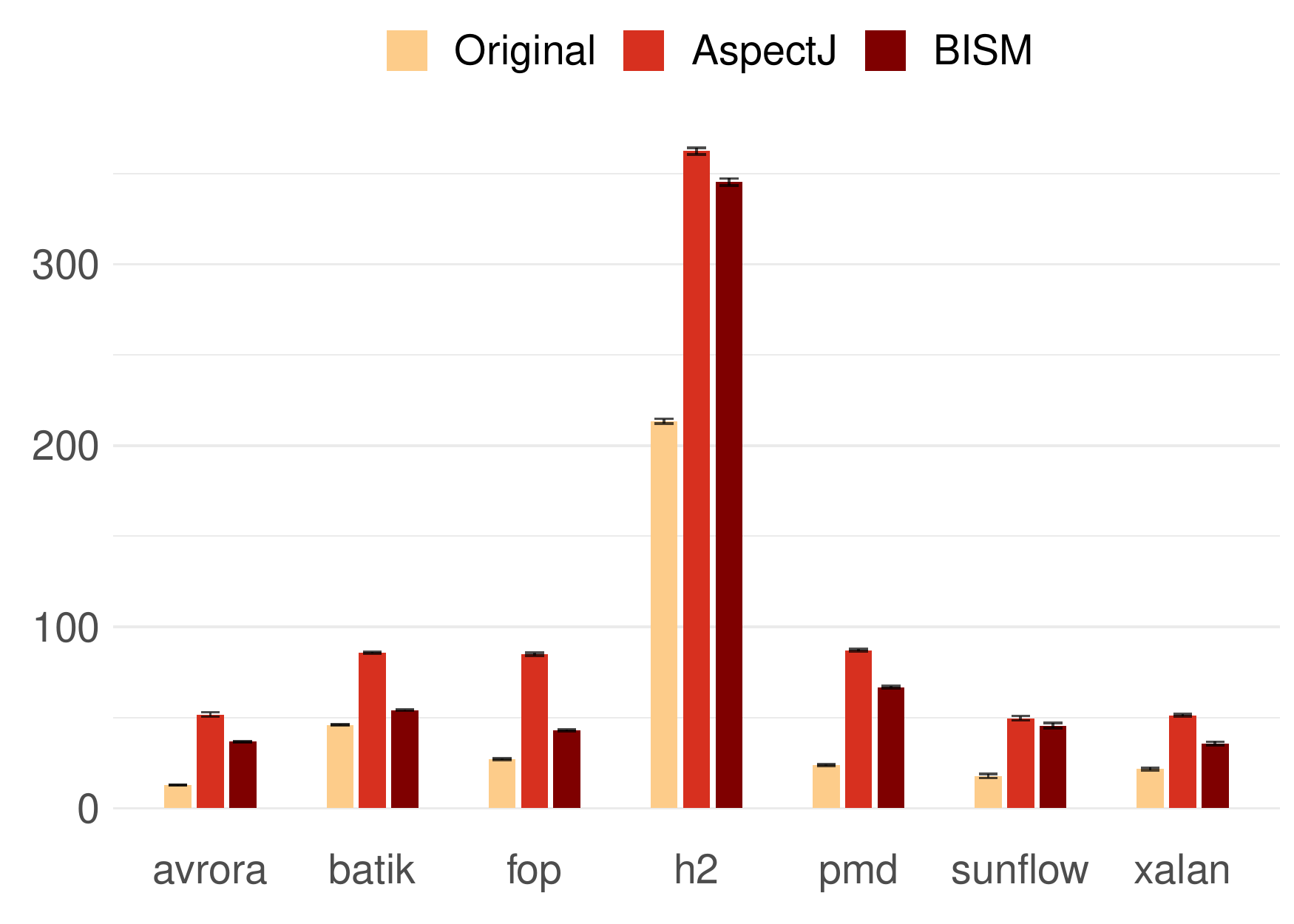}
    \caption{Used memory (\si{\mega\byte}).}
    \end{subfigure}
    \caption{DaCapo loadtime-time instrumentation.}
    \label{fig:dacapoloadtime}
    \end{figure}
\paragraph{Load-time evaluation.}
Figure~\ref{fig:dacapoloadtime} reports the runtime for the benchmarks.
BISM shows better performance over DiSL and AspectJ in all benchmarks. 
DiSL shows better performance than AspectJ except for the pmd benchmark.
For the pmd benchmark, this is mainly due to the fewer events emitted by AspectJ (see Table~\ref{tab:dacapo-size}).
We notice that AspectJ captures fewer events in benchmarks batik, fop, pmd, and sunflow.
This is due to its inability to instrument synthetic bridge methods generated by the compiler after type erasure in generic types.
BISM also shows less used-memory overhead over AspectJ in all benchmarks.
We mention that we did not measure the used memory for DiSL since it performs instrumentation on a separate JVM process.

\begin{figure}[htbp]
 \begin{subfigure}[b]{0.5\textwidth}
 \includegraphics[width=\textwidth]{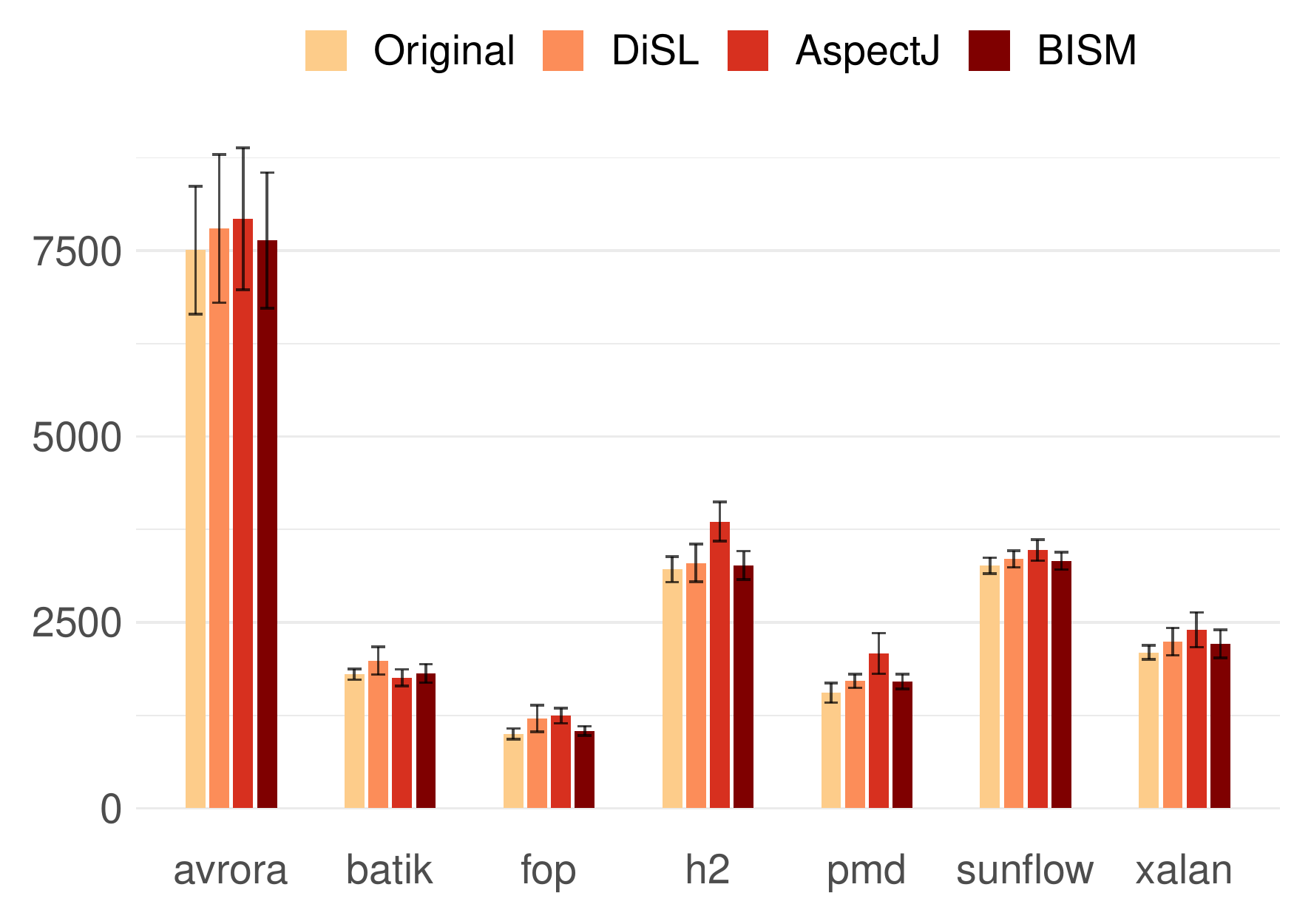}
 \caption{Runtime (ms).}
 \end{subfigure}
 \hfill
 \begin{subfigure}[b]{0.5\textwidth}
 \includegraphics[width=\textwidth]{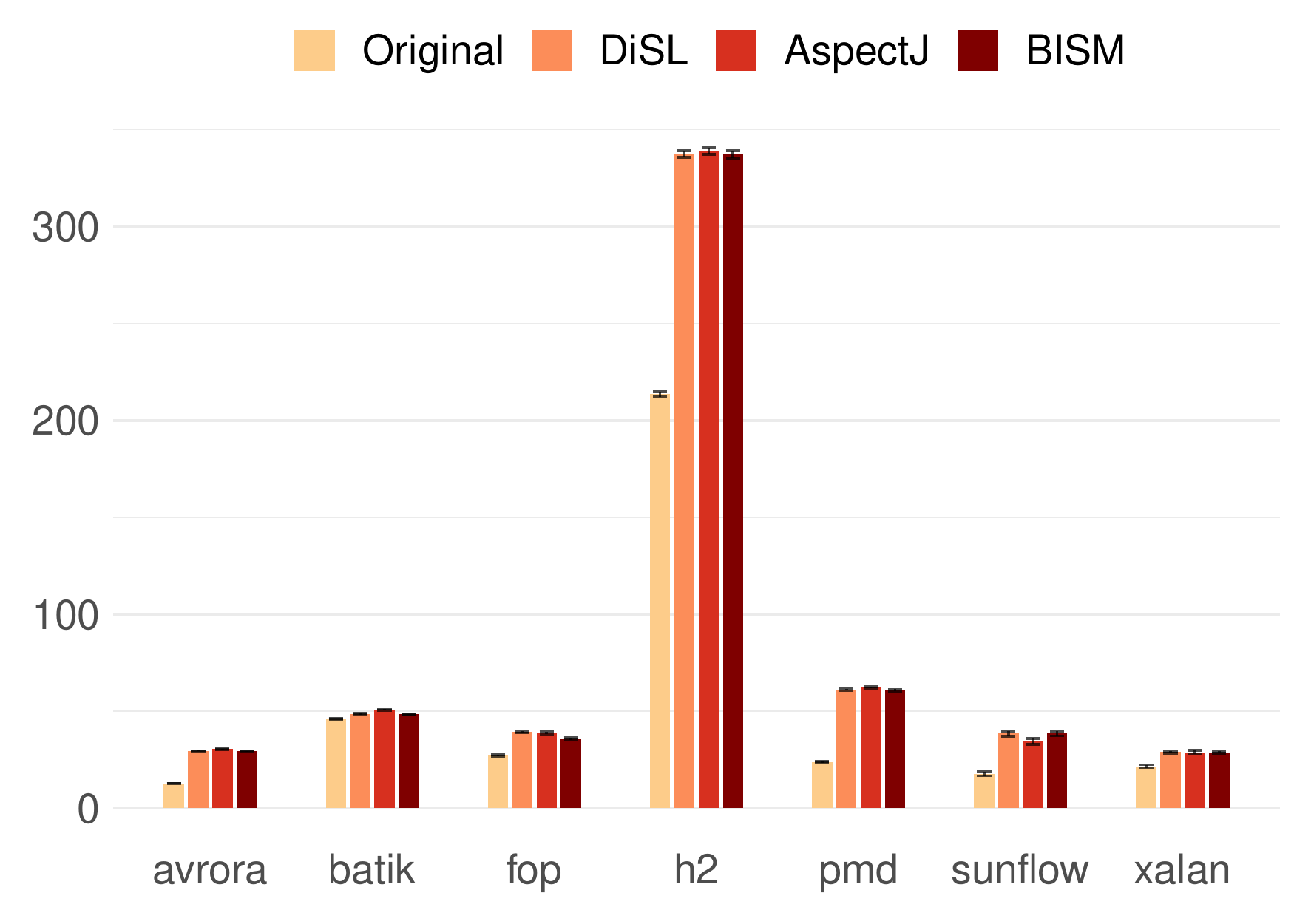}
 \caption{Used memory (\si{\mega\byte}).}
 \end{subfigure}
 \caption{DaCapo build-time instrumentation.}
 \label{fig:dacapobuildtime}
 \end{figure}

\paragraph{Build-time evaluation.}
We replace the original classes in the benchmarks with statically instrumented classes from each tool.
Figure~\ref{fig:dacapobuildtime} reports the runtime and used memory of the benchmarks.
%
%
BISM shows less runtime overhead in all benchmarks, except for batik where AspectJ emits fewer events.
BISM also shows less used-memory overhead, except for sunflow, where AspectJ emits much fewer events.

Table \ref{tab:dacapo-size} compares the instrumented bytecode. 
We report the number of classes in scope (Scope) and the instrumented (Instr.), and we measure the bytecode-size overhead (Ovh.) for each tool.
We also report the number of generated events after running the code (in millions). 
BISM and DiSL generate the same number of events, while Aspect (AJ) produces fewer events because of the reasons mentioned above.
The results show that BISM incurs less bytecode-size overhead for all benchmarks.
We notice that even with exception-handlers turned off, DiSL still wraps a targeted region with try-finally blocks when the \inlineJava{@After} annotation is used. 
This guarantees that an event is emitted after a method call, even if an exception is thrown.

\bgroup
\setlength{\tabcolsep}{5pt}
\begin{table*}[htbp]
 \centering
 \caption{For each benchmark in the DaCapo experiment, the table reports the number of classes in the scope of instrumentation (Scope), the instrumented classes (Instr.), the original (Ref.) and generated bytecode size and overhead per tool, and the number of events emitted, (\#) for BISM and DiSL, and AspectJ separately.}
 \renewcommand{\arraystretch}{1.5}
 \begin{tabular}{rrrrrrrrrrrrr}
 \toprule
 \multirow{2}*{\bf Benchmark} & \multicolumn{1}{c}{\bf Scope} & \multicolumn{1}{c}{\bf Instr.} & \multicolumn{1}{c}{\bf Ref.} & \multicolumn{2}{c}{\bf BISM} & \multicolumn{2}{c}{\bf DiSL} & \multicolumn{2}{c}{\bf AspectJ} & & \multicolumn{2}{c}{\bf Events (M)} \\
 & & & \si{\kilo\byte} & \si{\kilo\byte} & Ovh.\% & \si{\kilo\byte} & Ovh.\% & \si{\kilo\byte} & Ovh.\% & & \# & AspectJ \\
 \midrule
 avrora & 1,550 & 35 & 257 & 264 & 2.72 & 270 & 5.06 & 345& 34.24 & & 2.5 & 2.5 \\
 batik & 2,689 & 136 & 1,544 & 1,572 & 1.81 & 1,588 & 2.85 & 1,692 & 9.59 & & 0.5 & 0.4 \\
 fop & 1,336 & 172 & 1,784 & 1,808 & 1.35 & 1,876 & 5.16 & 2,267 & 27.07 & & 1.6 & 1.5 \\
 h2 & 472 & 61 & 694 & 704 & 1.44 & 720 & 3.75 & 956 & 37.75 & & 28 & 28 \\
 pmd & 721 & 90 & 756 & 774 & 2.38 & 794 & 5.03 & 980 & 29.63 & & 6.6 & 6.3 \\
 sunflow & 221 & 8 & 69 & 71 & 2.90 & 74 & 7.25 & 85 & 23.19 & & 3.9 & 2.6 \\
 xalan & 661 & 9 & 100 & 101 & 1.00 & 103 & 3.00 & 116 & 16.00 & & 1 & 1 \\
 \bottomrule
 \end{tabular}
 \label{tab:dacapo-size}
 \end{table*}
\egroup
\section{Related Work and Discussion} 
\label{sec:related-work}
Low-level code instrumentation is widely used for monitoring software and implementing dynamic analysis tools. 
Several instrumentation tools and frameworks in different programming languages have been developed. 
In the following, we compare BISM to other Java instrumentation tools. 
Nevertheless, there are several tools to instrument programs in different programming languages. 
For instance, to instrument C/C++ programs AspectC/C++~\cite{CoadyKFS01,Spinczyk05} (high-level) and LLVM~\cite{LattnerA04} (low-level) are widely used. 

\subsection{Instrumentation Tools}
Table~\ref{tab:tool-cmp} shows a comparison between some of the main tools for instrumenting Java programs among from which DiSL and AspectJ are the closest to BISM. 
The comparison considers some main features important for Java program (bytecode-) instrumentation, including bytecode visibility, the ability to insert bytecode instructions, and whether using the tool requires proficiency in bytecode, level of abstraction, and whether the instrumentation mechanism follows AOP Paradigm.

BCEL~\cite{BCEL} enables developers to perform static analysis and dynamically create and modify Java classes at runtime.
It is suitable for compilers, profilers, and bytecode optimization tools. 
Its API consists of a package for analyzing a Java class without having the source code, a package to generate or modify class objects dynamically, and tools to display the target class or convert it into HTML format or assembly language. 
BCEL does not follow the AOP Paradigm, and it requires proficiency in bytecode.

ASM~\cite{BrunetonASM02} is a bytecode manipulation framework utilized by several tools, including BISM.
ASM offers two APIs that can be used interchangeably to parse, load, and modify classes.
However, to use ASM, a developer must deal with the low-level details of bytecode instructions and the JVM.
BISM offers extended ASM compatibility and provides better abstraction with its aspect-oriented paradigm.

Javassist~\cite{Chiba00} is a class library for editing bytecodes in Java. It provides developers the ability to modify Java classes at runtime when being loaded by the JVM. 
Javassist provides two levels of API: source level and bytecode level. The source-level API does not require the developer to have knowledge about Java bytecode.

CGLIB~\cite{CGLIB} is a code generation library that allows developers to extend Java classes and add new classes at runtime. CGLIB makes use of ASM and some other tools. It provides some level of abstraction and can be used without having profound knowledge about bytecode.

DiSL~\cite{MarekVZABQ12} is a bytecode-level instrumentation framework designed for dynamic program analysis.
DiSL adopts an aspect-oriented paradigm. 
It provides an extensible joinpoint model by providing an extensible library for implementing custom pointcuts (\textit{markers}).
Even though BISM provides a fixed set of pointcuts (\locators{}), it performs static analysis on target programs 
to offer out-of-the-box additional and needed control-flow pointcuts with richer static context objects. 
Both tools do not offer dynamic pointcuts such as \textit{cflow}, \textit{this}, \textit{args} and \textit{if} from AspectJ.
As for dynamic context objects, both BISM and DiSL provide equal access. 
However, DiSL provides typed dynamic objects. 
Also, both tools are capable of inserting synthetic local variables. 
Both BISM and DiSL require basic knowledge about bytecode semantics from their users.
In DiSL, writing custom markers and context objects also requires additional ASM syntax knowledge.
However, DiSL does not allow the insertion of arbitrary bytecode instructions but provides a mechanism to write custom transformers in ASM that runs before instrumentation. 
Whereas BISM allows to directly insert bytecode instructions, as seen in \secref{sec:aes}.
Such a mechanism is essential in many runtime verification scenarios.
All in all, DiSL provides more features (mainly targeted for writing dynamic analysis tools) and enables dynamic dispatch amongst multiple instrumentations and analysis without interference~\cite{BinderMTA16}, while BISM is more lightweight, as shown by our evaluation.
Moreover, DiSL runs a separate virtual machine for instrumentation, while BISM runs as a standalone tool and requires no installation.

AspectJ~\cite{KiczalesHHKPG01} is the standard aspect-oriented programming~\cite{KiczalesLMMLLI97} framework highly adop\-ted for instrumenting Java applications. 
It provides a high-level language used in several domains like monitoring, debugging, and logging.
AspectJ provides a complex joinpoint model with an expressive pointcut expression language and dynamic pointcuts.
However, AspectJ cannot capture bytecode instructions and basic blocks joinpoints, making several instrumentation tasks impossible.
%
%
%
With BISM, developers can target single bytecode instructions and basic block levels, and they can access richer static joinpoint information. Moreover, BISM provides access to local variables and stack values.
Furthermore, AspectJ introduces a significant instrumentation overhead, as seen in \secref{sec:dacapo}, and provides less control on where instrumentation snippets get inlined.
In BISM, the advice methods are weaved with minimal bytecode instructions and are always inlined next to the targeted regions.

\bgroup
\setlength{\tabcolsep}{6pt}
\begin{table*}[]
 \centering
 \caption{
 Comparison of some of the main tools for instrumenting Java programs along some user-oriented features. 
 A checkmark (\dgreen\cmark) indicates that the tool provides the feature.
 A cross mark (\red\xmark) indicates that the tool does not provide the feature.
 A mixed checkmark/cross mark (\blue\hmark) indicates that the tool partially provides the feature.
 }
 \def\arraystretch{1.3}%
 \begin{tabular}{cccccccc}
 \toprule
 \multirow{2}*{\bf Feature} & \textbf{BCEL} & \textbf{ASM} & \textbf{Javassist} & \textbf{CGLIB} & \textbf{DiSL} & \textbf{AspectJ} & \textbf{BISM} \\
 & \cite{BCEL} & \cite{BrunetonASM02} & \cite{Chiba00} & \cite{CGLIB} & \cite{MarekVZABQ12} & \cite{KiczalesHHKPG01} & \cite{bism}\\
 \midrule
 Provides Bytecode Visibility & \dgreen\cmark & \dgreen\cmark & \dgreen\cmark & \dgreen\cmark & \dgreen\cmark & \red\xmark & \dgreen\cmark\\ 
 Allows Bytecode Insertion & \dgreen\cmark & \dgreen\cmark & \dgreen\cmark & \dgreen\cmark & \red\xmark & \red\xmark & \dgreen\cmark\\ 
 Requires no Bytecode Proficiency & \red\xmark & \red\xmark & \blue\hmark & \blue\hmark & \dgreen\cmark & \dgreen\cmark& \dgreen\cmark \\ 
 Provides high-Level Abstraction & \red\xmark & \blue\hmark & \blue\hmark & \blue\hmark & \dgreen\cmark & \dgreen\cmark& \dgreen\cmark \\ 
 Follows AOP Paradigm & \red\xmark & \red\xmark & \red\xmark & \red\xmark & \dgreen\cmark & \dgreen\cmark& \dgreen\cmark\\
 \bottomrule
 \end{tabular}
 \label{tab:tool-cmp}
\end{table*}
\egroup

\subsection{Transformer Composition}
%

Composition and interference problems in aspect-oriented programming have been studied in the literature.
Interference between different aspects is commonly addressed as \textit{aspect interactions} and \textit{aspect interference}.
The main objective is to detect places of interaction between different aspects (collision of transformers in BISM) and provide mechanisms to resolve conflicts.
In~\cite{Douence2002}, a framework for detection and resolution of aspect interactions is presented.
The work provides a formal model for aspect weaving and a framework for detecting and resolving conflicts between aspects using static analysis.
In~\cite{Takeyama2010}, the work focuses on unexpected behavior of combined advice (\text{advice interference}).
They show that controlling the order of execution of advice is not enough in some instances.
They propose an AspectJ extension with a new resolver \textit{around} advice for resolving interference where there is a conflict.
The introduced resolver can be implemented separately and composed to resolve interference between other resolvers.
BISM provides a built-in feature to capture transformer collision after a run.
However, we do not provide a mechanism for resolving conflicts, which can be addressed in our future work.

Also, the composition is studied concerning the base program and a single aspect.
In~\cite{Havinga2006} and \cite{Havinga2007}, composition conflicts related to introductions to the base program are modeled and detected using a graph-based approach.
Introductions are constructs that affect the structure of a class, such as changing the inheritance structure, adding and removing methods.
In BISM, such introductions are possible since the user is free to use the ASM structure and modify the class structure.
However, we do not address such conflicts and keep the user responsible for them. 
\section{Conclusions and Future Work}
\label{sec:conclusion}
%
\subsection{Conclusions}
\label{sec:future-works}
%
BISM is an effective tool for low-level and control-flow-aware instrumentation.
BISM is complementary to DiSL, which is better suited for dynamic analysis (e.g., profiling).
We demonstrate the versatility of BISM on several simple use cases.
Our first evaluation (\secref{sec:aes}) let us observe a significant advantage of BISM over DiSL due to BISM's ability to insert bytecode instructions directly, optimizing the instrumentation.
Our second and third evaluations (\secref{sec:transactions} and \secref{sec:dacapo}) confirm that BISM is a lightweight tool that can be used generally and efficiently in runtime verification.
We notice a similar bytecode performance between BISM and DiSL after build-time instrumentation since, in both tools, the instrumentation (monitor invocation) is always inlined next to the joinpoints. 
On the other hand, AspectJ advice is located in external classes, and the base program is instrumented to call these external classes at the joinpoints.

In load-time instrumentation, the gap between BISM and DiSL is smaller in benchmarks with many classes in scope and a small number of instrumented classes. 
This stems from the fact that BISM performs a complete analysis of the classes in scope to generate its static context. 
While DiSL generates static context only after marking the needed regions, which is more efficient.

Overall, we believe that BISM can be used as an alternative to AspectJ and DiSL for lightweight and expressive runtime verification and even runtime enforcement (cf.~\cite{Falcone10,FalconeMRS18,FalconeP19}) thanks to its bytecode insertion capability, equivalent or better performance, and ease of use.
The reported use cases also demonstrate BISM's versatility in providing support for static and dynamic analysis tools.
In addition to verification and enforcement, BISM provides easy access to program information and powerful modification primitives without requiring the source code.
%
\subsection{Future Work}
\label{sec:future-works}
%
We foresee several research avenues related to BISM, which can be split in two categories.

The first category relates to improvements of BISM itself.
We plan on extending the BISM language by adding more features to it, such as \locator{} guards that will facilitate the filtering of joinpoints to the user.
Guards can be annotations that decorate \locators{}.
They allow users to specify a filter on important static information such as scope, method signature, opcode for instruction, and others.
Also, the language can be expanded to add a declarative domain-specific language for specifying simple instrumentation directives.
This will allow users to write certain instrumentation specifications without the need to implement a custom transformer.
We will also add more advice methods to instance method invocations.

The second category relates to the use of BISM as a support for static and dynamic analysis.
As shown by our performance evaluation, BISM is more efficient than the long-used AspectJ instrumentation framework for runtime verification.
It is thus desirable to investigate the performance improvements that runtime verification tools could gain by using BISM as an alternative instrumentation tool.
Moreover, since BISM is capable of retrieving some static information about the program, static analysis tools can then be developed as transformers in BISM.
Such static analysis would not need the source code and execute using only the bytecode of the target programs.
Static analysis can also be beneficial in combination with a runtime verification approach to, e.g., enrich the information provided to a monitor or reduce the performance overhead of monitors.
For instance, BISM can be used to implement a control-flow-aware runtime verification tool.
Such an approach could (i) use both low-level control-flow events and higher-level events such as method calls and (ii) leverage some reachability analysis on the control flow.
Finally, using the bytecode insertion capabilities of BISM, effective runtime enforcement~\cite{FalconeMRS18,FalconeP19} tools can be implemented.



%
\newpage
\bibliographystyle{unsrt}
\bibliography{biblio}
\end{document}